\documentclass[lettersize,journal]{IEEEtran}
\usepackage{amsmath,amsfonts}
\usepackage{algorithmic}
\usepackage{algorithm}
\usepackage{array}
\usepackage[caption=false,font=normalsize,labelfont=sf,textfont=sf]{subfig}
\usepackage{textcomp}
\usepackage{stfloats}
\usepackage{url}
\usepackage{verbatim}
\usepackage{graphicx}
\usepackage{booktabs}
\usepackage{multirow}
\usepackage{xcolor}
\usepackage{pifont}
\usepackage[colorlinks, linkcolor=blue, anchorcolor=blue, citecolor=blue]{hyperref}
\usepackage{array, color, colortbl}
\begin{document}	
	\title{Toward Copyright Integrity and Verifiability via Multi-Bit Watermarking for Intelligent Transportation Systems}
	\author{Yihao Wang, Lingxiao Li, Yifan Tang, Ru Zhang, Jianyi Liu 
		\thanks{This work was accepted for publication in \textit{IEEE Transactions on Intelligent Transportation Systems}. This work was supported in part by the National Natural Science Foundation of China under Grant U21B2020 and in part by the BUPT Excellent Ph.D. Students Foundation under Grant CX2023120 and Grant CX20241055. (\textit{Corresponding author: Ru Zhang}.)}
		\thanks{Yihao Wang, Yifan Tang, Ru Zhang, and Jianyi Liu are with the School of Cyberspace Security, Beijing University of Posts and Telecommunications, Beijing 100876, China (e-mail: yh-wang@bupt.edu.cn; tyfcs@bupt.edu.cn; zhangru@bupt.edu.cn; liujy@bupt.edu.cn).}
		\thanks{Lingxiao Li is with the School of Cyberspace Security, Beijing University of Posts and Telecommunications, Beijing 100876, China, and also with the State Key Laboratory of Networking and Switching Technology, Beijing 100876, China (e-mail: lingxiao-li@bupt.edu.cn).}
		\thanks{Digital Object Identifier 10.1109/TITS.2025.3535932}}
	\markboth{IEEE Transactions on Intelligent Transportation Systems}%
	{Shell \MakeLowercase{\textit{et al.}}: A Sample Article Using IEEEtran.cls for IEEE Journals}
	
	
	\maketitle

	\begin{abstract}
Intelligent transportation systems (ITS) use advanced technologies such as artificial intelligence to significantly improve traffic flow management efficiency, and promote the intelligent development of the transportation industry. However, if the data in ITS is attacked, such as tampering or forgery, it will endanger public safety and cause social losses. Therefore, this paper proposes a watermarking that can verify the integrity of copyright in response to the needs of ITS, termed ITSmark. ITSmark focuses on functions such as extracting watermarks, verifying permission, and tracing tampered locations. The scheme uses the copyright information to build the multi-bit space and divides this space into multiple segments. These segments will be assigned to tokens. Thus, the next token is determined by its segment which contains the copyright. In this way, the obtained data contains the custom watermark. To ensure the authorization, key parameters are encrypted during copyright embedding to obtain cipher data. Only by possessing the correct cipher data and private key, can the user entirely extract the watermark. Experiments show that ITSmark surpasses baseline performances in data quality, extraction accuracy, and unforgeability. It also shows unique capabilities of permission verification and tampered location tracing, which ensures the security of extraction and the reliability of copyright verification. Furthermore, ITSmark can also customize the watermark embedding position and proportion according to user needs, making embedding more flexible.
	\end{abstract}

	\begin{IEEEkeywords}
		Intelligent transportation systems, Copyright, Watermark, Large language models, GenAI.
	\end{IEEEkeywords}

	\section{Introduction}\label{sec1}
	Intelligent transportation systems (ITS) \cite{tra1, itsappen1} use advanced information and communication technologies to optimize traffic flow, thereby improving road safety and reducing traffic accidents \cite{tra2}. The core goal of ITS is to achieve intelligent traffic management, including vehicle control and traffic safety management \cite{tra3, tra4}. With the application of generative artificial intelligence (GenAI) \cite{GPT42023, LLaMA32024, GPT22019, Gemini2023, ig1, itsappen3}, the quality and efficiency of ITS have been significantly improved \cite{trallm}. For example, data collection technology based on sensors and vehicle-to-vehicle communication enables traffic management systems to obtain large amounts of real-time data to conduct traffic flow prediction and road condition management. Although ITS has greatly improved the capabilities of traffic management and decision support, with the increasing number of smart devices and large-scale data analysis \cite{QIL, ig2}, data security and privacy issues \cite{trasec0, tbjz1, trasec01} have gradually become prominent. Malicious attackers may forge, tamper with, or steal data, which may cause the failure of the traffic management system and even cause traffic accidents \cite{trasec1, trasec2, trasec3}. In addition, how to ensure the source, copyright, and integrity of data in cross-departmental and cross-institutional data sharing and cooperation is also a major problem currently faced.
	
	In this context, watermarking \cite{KGW2023, NSW2023}, as a means of data protection, can effectively protect the copyright and source of data. It embeds invisible identification into the data without affecting the quality of data, ensuring that the data is not tampered with or abused during transmission and sharing. Especially in traffic forecasts, strategy analysis reports, cross-institutional cooperation data, etc., watermarkings are expected to provide effective copyright protection. This not only enhances the credibility of ITS data, but also provides guarantees for its legal use in multi-party cooperation \cite{REMARK-LLM2024}. 
	
	According to different goals, the current watermarking is generally categorized into “zero-bit” and “multi-bit” types \cite{WM_survey2024}. Zero-bit watermarking emphasizes the embedding of watermark signals, and the generated watermarked text is expected to be discriminated by detectors \cite{KGW2023, Adaptive2024}. Such schemes are usually studied around four aspects: “high quality” (embedding does not compromise data quality) \cite{KGW2023, NSW2023, He2024}, “easy to detect” (ensuring that the watermarked data can be easily discriminated) \cite{DiPmark2024, EWD2024}, “strong robustness” (the detector still can distinguish if the part of the data is tampered with) \cite{KGW-reliability2024, Unigram2024}, and “difficult to forge” (the watermarked data can resist forgery by attackers) \cite{Undetectable2024, Unforgeable2024}. Compared with zero-bit watermarking, there is less research on multi-bit watermarking. Multi-bit watermarking provides ideas for tracing the source of the text. It focuses more on the embedding of custom watermarks (such as ITS copyright information, timestamps, etc.) and extraction \cite{PRMW2024}. The essence of extraction is information restoration. Existing schemes have been studied around the following four aspects: “high quality” (embedding does not compromise data quality) \cite{CTWL2024}, “extractable” (the extractor can extract the watermark with a higher probability) \cite{CTWL2024}, {“strong robustness” \cite{PRMW2024}}, and “high payload” (high embedding rate, that is, the ratio of the amount of embedded watermark to the number of tokens is large) \cite{COLOR2024}. These works realize the embedding of custom copyright, greatly improving the flexibility and practical value of watermarks.
	
\begin{figure}[!htbp]
		\centering
		\includegraphics[width=0.99\linewidth]{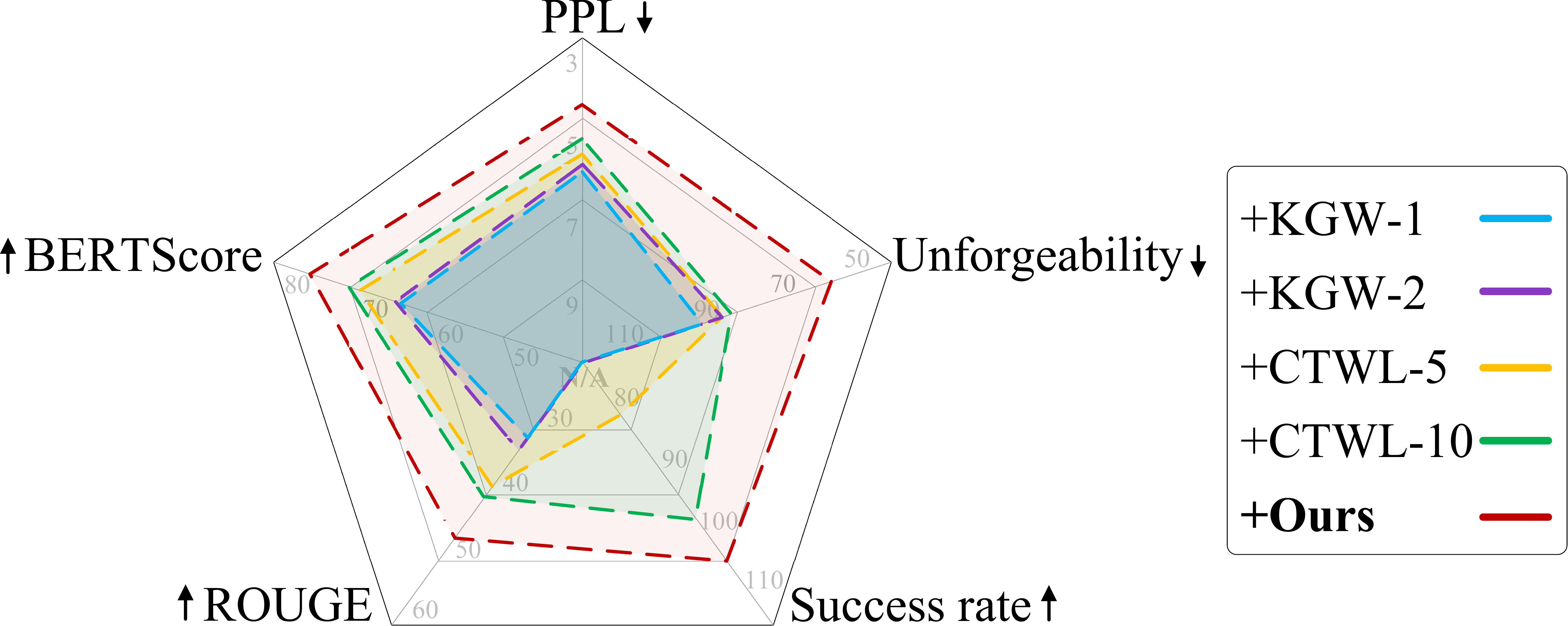}
		\caption{The performance of some SOTA schemes and the proposed ITSmark. “$\uparrow$” and “$\downarrow$” represent the higher / lower the value, the better the result. The metrics are found in “Section \ref{sec5}”.}\label{1}
	\end{figure}

	However, existing watermarking cannot meet the needs of ITS. First, the complete extraction of embedded watermark information cannot be guaranteed, resulting in insufficient reliability of data authenticity verification. Second, they lack a mechanism to verify the legitimacy of the verifier and cannot prevent unauthorized access and abuse of data. Finally, watermarks cannot effectively track the specific location of data tampering, resulting in the inability to timely identify and respond to data tampering or attacks. These issues directly affect the security, integrity, and credibility of data in ITS, and an improved watermark is urgently needed to address the core needs of ITS data protection, legitimacy verification, and tampering traceability. {The performance of some SOTA schemes is shown in Fig. \ref{1}.} Therefore, we introduce ITSmark (a water\textbf{mark}ing for \textbf{I}ntelligent \textbf{T}ransportation \textbf{S}ystems). This scheme focuses on extraction accuracy and extraction security while ensuring data quality and payload.

	Specifically, ITSmark can provide the following advanced functions for ITS: \textbf{(1) Custom and Entire Extraction}. ITSmark allows for the embedding of custom watermarks and ensures the entire extraction, protecting copyright integrity. During generation, the watermark space is built by copyright. Then, this space will be divided into multiple continuous bit segments based on logits and assigned them to all tokens in the alternative token list. Thus, the next token to be generated is determined by this watermark. \textbf{(2) Permission Verification}. ITSmark provides a permission verification mechanism. The extractor needs to have not only the watermarked data, but also the private key and cipher data of the sender. Only after the permission verification is passed can the extractor entirely extract the watermark. \textbf{(3) Traceability of Tampered Location}. If extraction is imprecise post-verification, it means that the content has been tampered with. ITSmark offers an efficient way to trace the tampered location precisely, providing a feasible method for forensic analysis. \textbf{(4) Unforgeability}. The watermarked data has strong unforgeability, enhancing the reliability of the watermark’s origin. \textbf{(5) Embedding Ratio}. ITSmark supports controlling the ratio of embedding, thus enhancing the flexibility of copyright protection.

	\section{Related Work}\label{sec2}
	\subsection{Intelligent Transportation Systems} 
	Intelligent transportation systems (ITS) have experienced rapid development in the past few decades \cite{RW1, RW2, RW3}. In recent years, the introduction of artificial intelligence and big data has greatly promoted the development of ITS. Traffic flow prediction \cite{RW6, RW7} based on big data analysis \cite{RW5} has become one of the core research directions of ITS. Researchers have developed efficient traffic flow prediction models by integrating sensor data and historical traffic data \cite{RW8, RW9, itsappen4}. These models can provide early warning of congestion and optimize traffic signal control. In addition, the development of autonomous driving technology has also become an important part of ITS. Vehicle-to-vehicle \cite{VV} and vehicle-to-infrastructure \cite{VI} communication technologies \cite{tbjz2, tbjz3} have enabled information sharing between vehicles, improving road safety and traffic efficiency. In particular, with the support of 6G networks \cite{RW2, itsappen2} and edge computing \cite{RW1, RW4}, real-time data processing \cite{RT, itsappen5} and low-latency communication \cite{LL} have enhanced the ITS intelligence level.
	
	At the same time, the application of generative artificial intelligence (GenAI) technology \cite{LLaMA22023, GLM42024} has brought new development opportunities to ITS. In terms of traffic simulation and strategy generation, GenAI can generate simulated data in complex traffic scenarios for optimizing traffic management strategies \cite{tra3, tra4}. This technology effectively solves the limitations of traditional methods in dealing with sparse data and abnormal scenarios \cite{TRS1, TRS2}. In addition, researchers have also explored the application of LLMs in the automatic generation of traffic reports \cite{TRS3}, further improving the efficiency and automation of ITS.

	\subsection{Watermarkings} 
	Zero-bit watermarking can embed watermarks by modifying the logits or adjusting token sampling. Regarding the modification of logits, Kirchenbauer et al. \cite{KGW2023} first proposed a watermarking, which generates a green-red list using a hash of the previous token. The logits of tokens in the green list are increased, while those in the red list are decreased. Subsequent improvements to the \cite{KGW2023} scheme by Takezawa et al. \cite{NSW2023} and Fu et al. \cite{Fu2024} enhanced the quality of the generated watermarked text. Lee et al. \cite{SWEET2024} believed that high-entropy logits indicate token uncertainty. Multi-bit watermarking embeds the watermark by semantic \cite{AWT2021}, format \cite{Unicode2016}, and syntactic \cite{Robust_Multi-bit2023} modifications and replacement \cite{Context-aware2022}. Wang et al. \cite{CTWL2024} used a proxy language model to guide vocabulary division, making the probabilities of available and unavailable parts close to the same. TABLE \ref{rw} summarizes the functions for ITS of some representative watermark works.

	\begin{table}[!htbp]
		\centering
		\setlength{\tabcolsep}{1.15mm}
		\caption{The functions for ITS of some representative watermark works.}\label{rw}%
		\begin{tabular}{c||ccccccc|c}
			\toprule[1.2pt]
			\multirow{2}[1]{*}{Functions} &\multicolumn{8}{c}{Schemes}\\
			\cmidrule{2-9}
			& \cite{KGW2023}  & \cite{Adaptive2024}  &\cite{He2024}& \cite{Undetectable2024}  & \cite{PRMW2024}&\cite{CTWL2024} & \cite{COLOR2024}  & \textbf{Ours} \\
			\midrule[0.5pt]
			Quality & \ding{51}  &  \ding{51} &\ding{51}&  \ding{51}  &\ding{51}& \ding{51} &  \ding{51} & \ding{51} \\
			Custom watermark & \ding{55} & \ding{55}  &\ding{55}& \ding{55}  & \ding{51}& \ding{51}  & \ding{51}  & \ding{51} \\
			High payload&     \ding{55}  & \ding{55}&     \ding{55} &     \ding{55}  &\ding{51}&     \ding{51}  &  \ding{51}     & \ding{51} \\
			Entire extraction &   \ding{55}    &   \ding{55}    &\ding{55}&   \ding{55}    &\ding{55}&     \ding{55}  &    \ding{55}   &  \ding{51}\\
			Verification &  \ding{55}   & \ding{55} & \ding{55} &    \ding{55}   &\ding{55}&   \ding{55}    &    \ding{55}   & \ding{51} \\
			Tampering traceability & \ding{55} &    \ding{55} &   \ding{55}    &\ding{55}&    \ding{55}   &   \ding{55}    &    \ding{55}   &  \ding{51}\\
			Unforgeability &    \ding{55}    &  \ding{55}  &   \ding{55}  &     \ding{51}  &\ding{55}&      \ding{55} &     \ding{55}  & \ding{51} \\
			Partial embedding &     \ding{55}  & \ding{55}&     \ding{55} &     \ding{55}  &\ding{55}&     \ding{55}  &  \ding{55}     & \ding{51} \\
			\bottomrule[1.2pt]
		\end{tabular}%
		\vspace{-2.5ex}
	\end{table}%

	\section{ITSmark Methodology}\label{sec3}
	\subsection{Overall}
\begin{figure}[!htbp]
		\centering
		\includegraphics[width=0.85\linewidth]{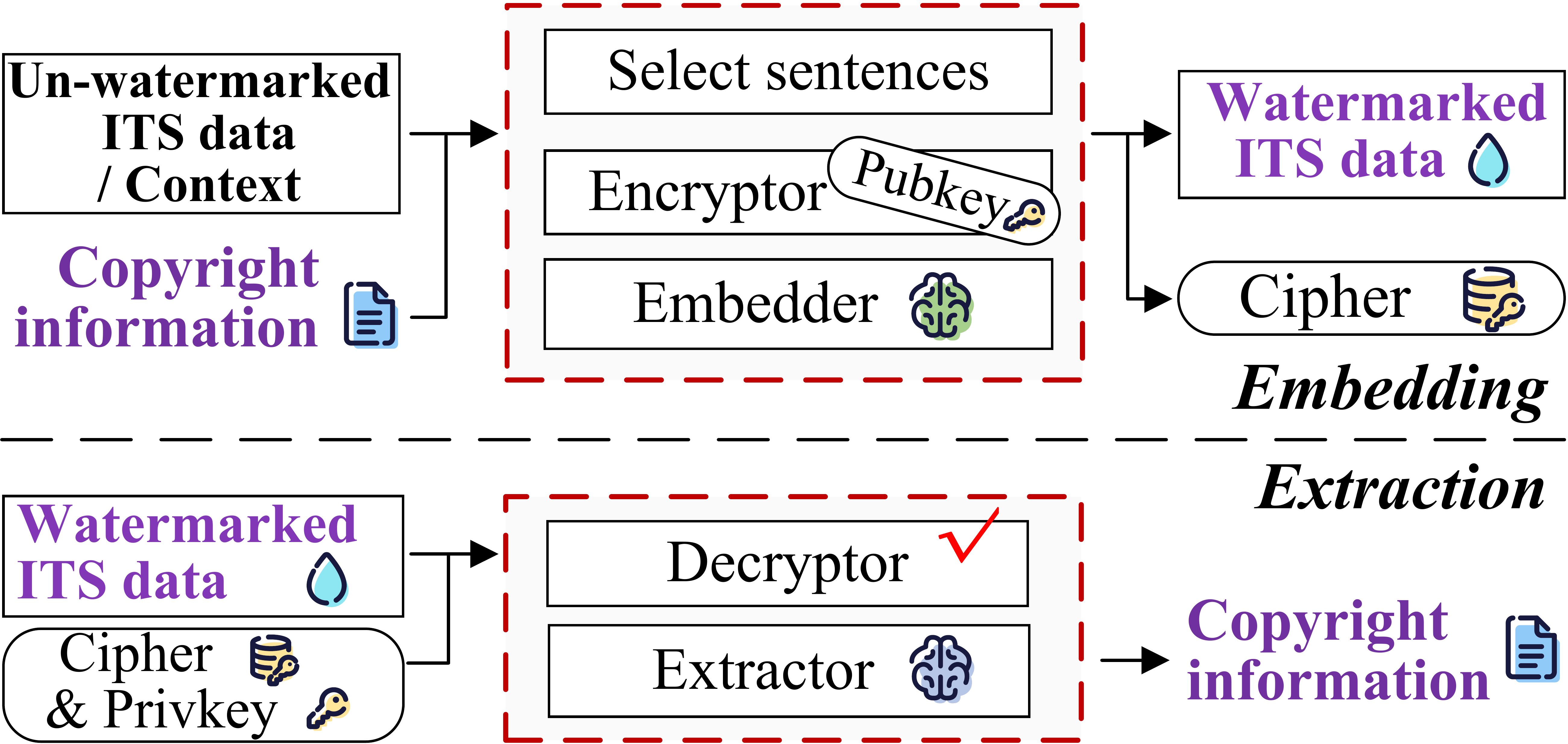}
		\caption{The embedding and extraction processes of ITSmark. “Pubkey” and “Privkey” are the public key and the private key. They are generated by the encryption algorithm. {{The encryption algorithm can be arbitrary. When it is actually deployed, an encryption algorithm with efficient key management and distribution can be adopted, such as PKI that can effectively manage the distribution and verification of public keys or key lifecycle management tools to help automate key generation, distribution, update, revocation, and other operations.}} Before extraction, permission verification is necessary. Only after successful verification can the copyright be extracted.}\label{fig1}
	\end{figure}

	As demand for copyright protection increases, watermarking also needs to provide advanced functionalities such as copyright integrity verification and permissions. That is, the designed scheme should not only ensure the data quality, but also support the embedding of custom watermarks and the security and entire of watermark extraction. Here, To meet these needs, this paper proposes the ITSmark, and its embedding and extraction processes are shown in Fig. \ref{fig1}.

	\begin{figure*}[!b]
		\centering
		\includegraphics[width=0.85\linewidth]{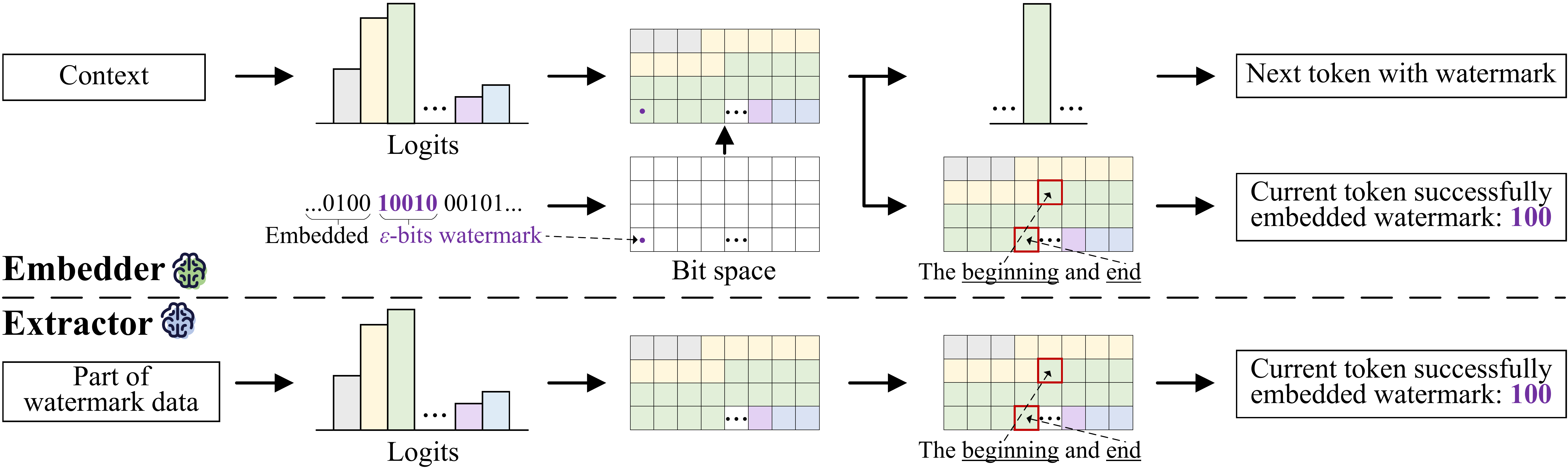}
		\caption{The working principle of the “Embedder” and “Extractor” in ITSmark. Here only a certain moment's embedding and extraction processes are shown. The context is input into the “Embedder” to obtain the next token containing the watermark information.}\label{fig2}
	\end{figure*}

	The expressions for “Embedding” and “Extraction” are:
	
	\begin{equation}\label{eq1}
	\left\{ {\begin{array}{*{1}{r}}
		Embedding:\mathcal{O} \times \mathcal{M} \times \mathcal{PK} \to \mathcal{W},\mathcal{C},\ \ Em(o,\\ m,pubkey) = w,cipher\\
		Extraction:\mathcal{W} \times \mathcal{C} \times \mathcal{SK} \to \mathcal{M},\ \ Ex(w,\\ cipher,privkey) = m
		\end{array}} \right..
	\end{equation}
	
	\noindent When embedding, $Em( \cdot )$ uses the input $o \in \mathcal{O}$ (such as Context), multi-bit watermark information $m \in \mathcal{M}$, and public key $pubkey \in \mathcal{PK}$ to obtain the watermarked ITS data $w \in \mathcal{W}$ and cipher data $cipher \in \mathcal{C}$. When extracting, $Ex( \cdot )$ uses the watermarked ITS data $w$, cipher data $cipher$ and the private key $prikey \in \mathcal{SK}$ corresponding to $pubkey$. If either the private key or cipher data is incorrect, the verification fails, and the watermark cannot be entirely extracted.
	
	The working principle of the “Embedder” and “Extractor” in Fig. \ref{fig1} is shown in Fig. \ref{fig2}.

	\subsection{Next token determined by watermark information} 
	Given a context or an unwatermarked data (token sequence) of length $t-1$ ${x_{1:t - 1}} = ({x_1},{x_2}, \cdots ,{x_{t - 1}})$. Input ${x_{1:t - 1}}$ into LLMs to calculate the conditional probabilities, i.e., the logits $P({v_i}|{x_{1:t - 1}}),v_i \in \mathcal{V}, {\rm{ }}i \in [1,n], \mathcal{V}(|\mathcal{V}| = n)$. After obtaining the logits, ITSmark determines the next token by $\mathcal{M}$ to embed the watermark. Given a $\mathcal{M}$, which can be converted into a binary $a$-bit ${m_{1:a}} = ({m_1}, \cdots ,{m_a})$ using ASCII or other encodings. ${m_{1:a}}$ is not embedded all at once, but during the generation process, $\varepsilon$ bits are read sequentially to process $\varepsilon$-bit watermark ${m_{1:\varepsilon }} = ({m_1}, \cdots ,{m_\varepsilon }),{\rm{ }}\varepsilon  \le a$. ${m_{1:\varepsilon }}$ builds the multi-bit space $\mathcal{B}, \ {\rm{ }}\left|\mathcal{B}\right| = {2^\varepsilon }$, and ${m_{1:\varepsilon }}$ certainly exist in $\mathcal{B}$. 
	
	Then, we divide $\mathcal{B}$ into multiple consecutive bit segments $\mathcal{S}$. A bit segment is assigned to a token continuously. To ensure the quality, we only assign $\mathcal{S}$ to all tokens. So, how to divide the $\mathcal{B}$ plays a key role in determining the next token. Assume that the $l$-th to $j$-th consecutive bits being assigned to the $i$-th token $p(l,j,i)$ is determined by the logit of the token, and is:
	
	\begin{equation}\label{eq4}
	p(l,j,i) = \frac{{P{{({v_i}|{x_{1:t - 1}})}^\lambda }}}{{\sum\nolimits_{{v_i} \in {\mathcal{V}_{\mathop{\rm}\nolimits}}} {P{{({v_i}|{x_{1:t - 1}})}^\lambda }} }},
	\end{equation}
	
	\noindent where, $\lambda$ is the weight parameter. By processing with the additional condition, each token in $\mathcal{V}_{\mathop{\rm green}\nolimits}$ will get some continuous bits. And it will be weighted and normalized by formula (\ref{eq4}). Correspondingly, the number of bits obtained by all tokens in the green list is distributed as ${\bf{p}}(\lambda ) = \left[p(l,j,1), \cdots ,p(l,j,i), \cdots \right]$.
	
	Here we want to explore how the overall distribution ${\bf{p}}(\lambda )$ changes with the change of $\lambda$. So, we take it as a continuous function and take its derivative. Assume that the expectation of ${\bf{p}}(\lambda)$ is $\mathbb{E}[{\bf{p}}(\lambda)]$, then the variance is ${\mathop{\rm Var}\nolimits} ({\bf{p}}(\lambda )) = \mathbb{E}[{\bf{p}}{(\lambda )^2}] - {(\mathbb{E}[{\bf{p}}(\lambda )])^2}$. We take the derivative of ${\mathop{\rm Var}\nolimits} ({\bf{p}}(\lambda ))$:
	
	\begin{equation}\label{eq5}
	\frac{d}{{d\lambda }}{\mathop{\rm Var}\nolimits} ({\bf{p}}(\lambda )) = \frac{d}{{d\lambda }}\mathbb{E}[{\bf{p}}{(\lambda )^2}] - \frac{d}{{d\lambda }}{(\mathbb{E}[{\bf{p}}(\lambda )])^2}.
	\end{equation}
	
	\noindent $\mathbb{E}[{\bf{p}}(\lambda)^2]$ is:
	
	\begin{equation}
	{\mathbb{E}[{\bf{p}}(\lambda)^2]}=\sum\nolimits_i {{p_i}{{(\lambda )}^2}},
	\end{equation}
	
	\noindent where, ${p_i}(\lambda ) = {\textstyle{{P{{({v_i}|{x_{1:t - 1}})}^\lambda }} \over {Z(\lambda )}}}$, and $Z(\lambda )$ is:
	
	\begin{equation}
	Z(\lambda ) = \sum\nolimits_j {P{{({v_j}|{x_{1:t - 1}})}^\lambda }},
	\end{equation}
	
	\noindent the derivative of ${p_i}{(\lambda )^2}$ is calculated:
	
	\begin{equation}
	\frac{d}{{d\lambda }}{p_i}{(\lambda )^2} = 2{p_i}(\lambda )\frac{d}{{d\lambda }}{p_i}(\lambda ).
	\end{equation}
	
	\noindent Next, we calculate the derivative of ${p_i}(\lambda)$ with respect to $\lambda$:
	
	\begin{equation}
	\begin{array}{c}
	\frac{d}{{d\lambda }}{p_i}(\lambda ) = {p_i}(\lambda ) \times \\ \left( {{\mathop{\rm In}\nolimits} P({v_i}|{x_{1:t - 1}}) - \sum\limits_j {{p_j}(\lambda ){\mathop{\rm In}\nolimits} P({v_j}|{x_{1:t - 1}})} } \right)
	\end{array},	
	\end{equation}
	
	\noindent so:
	
	\begin{equation}
	\begin{array}{c}
	\frac{d}{{d\lambda }}{p_i}{(\lambda )^2} = 2{p_i}{(\lambda )^2} \times \\ \left( {{\mathop{\rm In}\nolimits} P({v_i}|{x_{1:t - 1}}) - \sum\limits_j {{p_j}(\lambda ){\mathop{\rm In}\nolimits} P({v_j}|{x_{1:t - 1}})} } \right)
	\end{array}.
	\end{equation}
	
	\noindent Adding up all the terms of $i$, we get:
	
	\begin{equation}
	\begin{array}{c}
	\frac{d}{{d\lambda }}\mathbb{E}[{\bf{p}}{(\lambda )^2}]= 2\sum\limits_i {{p_i}{{(\lambda )}^2}} \times \\ \left( {{\mathop{\rm In}\nolimits} P({v_i}|{x_{1:t - 1}}) - \sum\limits_j {{p_j}(\lambda ){\mathop{\rm In}\nolimits} P({v_j}|{x_{1:t - 1}})} } \right)
	\end{array},
	\end{equation}
	
	\noindent due to $\mathbb{E}[{\bf{p}}(\lambda )] = \sum\nolimits_i {{p_i}(\lambda )}  = 1$, so:
	
	\begin{equation}
	{(\mathbb{E}[{\bf{p}}(\lambda )])^2} = 1,
	\end{equation}
	
	\noindent its derivative is zero: 
	
	\begin{equation}
	\frac{d}{{d\lambda }}{(\mathbb{E}[{\bf{p}}(\lambda )])^2}=0.
	\end{equation}
	
	\noindent According to the definition of variance: 
	
	\begin{equation}
	{\mathop{\rm Var}\nolimits} ({\bf{p}}(\lambda )) = \mathbb{E}[{\bf{p}}{(\lambda )^2}] - {(\mathbb{E}[{\bf{p}}(\lambda )])^2},
	\end{equation}
	
	\noindent so its derivative is:
	
	\begin{equation}
	\begin{array}{c}
	\frac{d}{{d\lambda }}{\mathop{\rm Var}\nolimits} ({\bf{p}}(\lambda ))= 2\sum\limits_i {{p_i}{{(\lambda )}^2}} \times \\ \left( {{\mathop{\rm In}\nolimits} P({v_i}|{x_{1:t - 1}}) - \sum\limits_j {{p_j}(\lambda ){\mathop{\rm In}\nolimits} P({v_j}|{x_{1:t - 1}})} } \right)
	\end{array}.
	\end{equation}
	
	\noindent Since ${p_i}{(\lambda )^2}$ is non-negative, and larger ${p_i}{(\lambda )}$ corresponds to larger $P({v_j}|{x_{1:t - 1}})$, so this deviation is non-negative. Therefore, when $\lambda>0$, ${\textstyle{d \over {d\lambda }}}{\mathop{\rm Var}\nolimits} ({\bf{p}}(\lambda )) > 0$, this means that as $\lambda$ decreases, the ${\mathop{\rm Var}\nolimits} ({\bf{p}}(\lambda ))$ will decrease and ${\bf{p}}(\lambda )$ will become smooth. In other words, the larger $\lambda$, the more priority is placed on tokens with larger logits, and these tokens will get most of the multi-bit space. When $\lambda=0$, each token will have the same priority in the number of bits, so that the number of bits finally assigned is also the same. In addition, due to the diversity of the watermark $\mathcal{M}$, the distribution of the corresponding binary ${m_{1:a}}$ conforms to the “Bernoulli distribution” with the same probability of 0 and 1 appearing. To ensure the quality, ${m_{1:\varepsilon}}$ needs to be more likely to fall within the larger bit range of $P({v_i}|{x_{1:t - 1}})$. 
	
	Therefore, $\lambda$ cannot be too small, otherwise, the probability of selecting tokens with smaller logits will increase, which is not conducive to the quality. And $\lambda$ also cannot be too large, otherwise, the token with maximum logit will be selected each time, which is not conducive to the embedding. For experiments on $\lambda$ and $\varepsilon$, please see Section \ref{sec542}.
	
	In this way, the multi-bit space $\mathcal{B}$ is divided into multiple bit segments $\mathcal{S}$ and assigned these segments completely to ${v_i} \in {\mathcal{V}_{\mathop{\rm}\nolimits}}$. At this time, ${m_{1:\varepsilon }}$ certainly fall within a bit segment $s \in \mathcal{S}$. For example, ${m_{1:\varepsilon }}=10010$, $s: [10000, 10011]$, ${m_{1:\varepsilon }} \in s$. This segment $s$ corresponding to a certain token is to be generated. Thus, this token contains part of $\mathcal{M}$.

	\subsection{The watermark embedded at a moment} 
	To entirely extract the currently embedded watermark, the process of “Extractor” and “Embedder” needs to be completely reversible. However, the token corresponds to a bit segment $s$, which contains a continuous range of bits. Any bits $m'$, which $m' \in s$, will determine this token. So during embedding, the relationship between the watermark read and the next token is the many-to-one. During the extraction process, the relationship between the next token and the watermark is one-to-many. Even if the next token in the watermarked text is known, in this situation, any watermark bits belonging to this $s$ will be regarded as the “potential watermark information”. Therefore, the many-to-one relationship needs to be converted into a one-to-one during embedding. We perform bit string matching between ${m_{1:\varepsilon}}$ and the beginning $\beta$ and end $\beta'$ of $s=[\beta ,\beta ']$ that determines the token and have:
	
	\begin{equation}\label{eq7}
	\begin{array}{r}
	{m_{1:\varepsilon }} \in [\beta ,\beta '] \Rightarrow \exists p,{\rm{ }}{\mathop{\rm s}\nolimits} .t.{\rm{ }}({\beta _1}, \cdots ,{\beta _p}) = ({m_1}, \\ \cdots ,{m_p}) = ({\beta _1}', \cdots ,{\beta _p}'),p \le \varepsilon ,{\rm{ }}p \in \mathbb{Z}
	\end{array}.
	\end{equation}
	
	\noindent Simplifying formula (\ref{eq7}):
	
	\begin{equation}\label{eq8}
	\exists p,{\rm{ }}{\mathop{\rm s}\nolimits} .t.{\rm{ }}({\beta _1}, \cdots ,{\beta _p}) = ({\beta _1}', \cdots ,{\beta _p}'),p \le \varepsilon .
	\end{equation}
	
	\noindent i.e. $s: [10000, 10011]$, the watermark embedded at this moment is $100$. When $p = 0$, their match is an empty string. That is, if the bit segment is determined, the match of the beginning and end of the segment is unique and must be a substring ${m_{1:p}}$ of ${m_{1:\varepsilon }}$. The information successfully embedded in the current token is not the read ${m_{1:\varepsilon }}$, but ${m_{1:p}}$, the same first $p$ bits of the beginning and end of the corresponding segment of ${m_{1:\varepsilon}}$. Therefore, only the first $p$ bits of $s$ are embedded, the next moment, which should be read is ${m_{p + 1:p + 1 + \varepsilon }}$.

	\subsection{Watermark extracted} 
	The process of “Extractor” and “Embedder” is inverse, and extraction needs to use the same system as embedding. That is the same LLMs and division. The current watermark extraction process is shown in Algorithm \ref{Alg1}.

	\begin{algorithm}[!htbp]
		\small
		\caption{\small Extraction of the current embedded watermark.} 
		\label{Alg1}
		\begin{algorithmic}[1]
			{
				\REQUIRE The first part of the watermarked data.
				\STATE Calculate $P({v_i}|{x_{1:t - 1}}),{v_i} \in \mathcal{V}$ of ${x_{1:t - 1}}$;
				\STATE Divide the multi-bits space by logits of ${v_i} \in {\mathcal{V}}$, and assign $\mathcal{S} \Rightarrow {v_i}, v_i \in \mathcal{V} $;
				\STATE Read the $\varepsilon$-bit watermark ${m_{1:\varepsilon }}$ and determine $v_i$;
			}
			\STATE Locate the bit segment $s=[\beta ,\beta ']$ according to the next token in the watermarked data;
			\STATE Match the beginning and end of the segment $s$, $\exists p, \ {\mathop{\rm s}\nolimits} .t. \ ({\beta _1}, \cdots ,{\beta _p}) = ({\beta _1}', \cdots ,{\beta _p}'), \ p \le \varepsilon$;
			\ENSURE The currently embedded ${m_{1:p}} = ({\beta _1}, \cdots ,{\beta _p})$.
		\end{algorithmic} 
	\end{algorithm}

	\subsection{Embedding Ratio}
	ITSmark can control the position and proportion of embedded watermarks, that is, you can choose to embed watermarks in certain sentences (“Partial embedding”) or in the entire data (“Full embedding”). In the embedding part of Fig. \ref{fig1}, we present the “Select sentences” module. 
	
	Intuitively, “Partial embedding” is more suitable for fewer watermarks and higher requirements on the quality and content precise of the watermarked text. Initially, it selects the sentence to embed the watermark according to certain rules (such as entropy and significance). Next, for the first selected sentence, its corresponding context is used to regenerate this sentence so that it contains the watermark. When generating the second watermarked sentence, the context used is the updated text containing the first watermarked sentence generated previously. And so on, by gradually replacing the original sentences, the final watermarked text is obtained. The cipher data contains the context, $\lambda$, and $\varepsilon$, but also contains the sequence of watermarked sentences “$sentence$ $list$”. For experiments on embedding ratio $\eta$, please see Section \ref{sec541}.
	
	\vspace{-1ex}
	\subsection{Permission Verification}
	If the watermark can be entirely extracted only by using the watermarked data, it will lead to the leakage and abuse of copyright information. Therefore, ITSmark needs to perform verification before extraction, which can ensure that only legally authorized users can extract the watermark.

	\setcounter{table}{2}
	\begin{table*}[b]
		\centering
		\vspace{-3ex}
		\caption{{Overview of each part of the experiments.}}\label{tab2}
		\begin{tabular}{cc||cccc}
			\toprule[1.2pt]
			\multicolumn{2}{c||}{Section} & Focus & Embedding ratio& TABLE & Fig. \\
			\midrule
			\multirow{3}{*}{Comparison study} & \ref{sec521} & Quality & Full &\ref{tabb1}& \ref{fig4}, \ref{fig5}\\
			& \ref{sec522} & Embedding time \& Extraction success rate & Full& \ref{tab3} & -- \\
			& \ref{sec523} & Unforgeability & Full&-- & \ref{fig6} \\
			\midrule
			\multirow{4}{*}{ITSmark exploration} & \ref{sec531} & Permission verification & Full&\ref{tabe1}& \ref{fig7} \\
			& \ref{sec532} & Traceability of tampered locations & Full& \ref{taba}& \ref{fig8} \\
			& \ref{sec541} & Ablation on embedding ratios $\eta$ & Partial &\ref{tabb2}, \ref{tabf1} &\ref{fig9}, \ref{fig10} \\
			& \ref{sec542} & Ablation on weight $\lambda$ and number of read bits $\varepsilon$ & Full&\ref{tab5}& \ref{fig12} \\
			\bottomrule[1.2pt]
		\end{tabular}%
	\end{table*}%

	When embedding, ITSmark will generate a pair of public and private keys $pubkey$ and $privkey$ by the encryption algorithm. The $pubkey$ is used to encrypt data such as Prompt, $\lambda$, and $\varepsilon$ to obtain the cipher data $cipher$. The extraction process not only needs the watermarked text, but also needs to have this $cipher$ and the correct $privkey$ to decrypt $cipher$. In this way, the permission verification can be passed, and the multi-bit watermark can be entirely extracted. For the experiment of permission verification, please see Section \ref{sec531}.

	\section{Experiments}\label{sec5}
	\subsection{Settings}
	\setcounter{table}{1}
	\subsubsection{Dataset and model configuration} Commonly used datasets in ITS and NLP are used to evaluate the performance of ITSmark, which are: TV \cite{KDD2023}, BDD \cite{KDD2023}, AlpacaFarm \cite{INT2023}, and FinQA \cite{FinQA2018}. The LLMs of LLaMA2-7B \cite{LLaMA22023} and ChatGLM3-6B \cite{GLM42024} are selected, where the generation strategies are Top\_k \textbf{Sample} and Beam \textbf{Search}. Some data are generated for experiments as shown in TABLE \ref{tab1}.

	\begin{table}[!htbp]
		\centering
		\footnotesize
		\vspace{-2ex}
		\setlength{\tabcolsep}{1.1mm}
		\caption{Detailed information of ITSmark data.}\label{tab1}
		\begin{tabular}{c||cccc|c}
			\toprule[1.2pt]
			\multirow{2}[2]{*}{\textbf{LLaMA2-7B}} & \multicolumn{4}{c|}{Num of tokens (Avg.)} & \multicolumn{1}{c}{\multirow{2}[2]{*}{Total}} \\
			\cmidrule{2-5}&TV&BDD& AlpacaFarm  & FinQA &  \\
			\midrule
			\cellcolor{gray!15}Sample &\cellcolor{gray!15}69.57&\cellcolor{gray!15}124.14& \cellcolor{gray!15}182.82 &  \cellcolor{gray!15}165.84 &  \\
			+Ours  &67.19&121.26& 189.77 & 166.21 &  \\
			\cellcolor{gray!15}Search &\cellcolor{gray!15}68.43&\cellcolor{gray!15}126.81& \cellcolor{gray!15}183.40 &  \cellcolor{gray!15}171.48 & ITSmark data: \\
			+Ours  &67.50&127.72& 181.57 &167.91& 25,607 \\
			\cmidrule{1-5}    \textbf{ChatGLM3-6B} &TV&BDD& AlpacaFarm & FinQA &  \\
			\cmidrule{1-5}    
			\cellcolor{gray!15}Sample &\cellcolor{gray!15}79.56&\cellcolor{gray!15}87.98& \cellcolor{gray!15}116.83 & \cellcolor{gray!15}104.39 & ITSmark tokens:\\
			+Ours  &80.14&92.05& 114.46 & 97.75 & 3,591,714\\
			\cellcolor{gray!15}Search &\cellcolor{gray!15}92.86&\cellcolor{gray!15}89.96& \cellcolor{gray!15}120.54 & \cellcolor{gray!15}100.96 &  \\
			+Ours  &90.23&86.18& 119.46 & 103.34 &  \\
			\bottomrule[1.2pt]
		\end{tabular}%
	\end{table}%

	\subsubsection{Baselines} We selected recent watermarking for comparison: zero-bit KGW-1/KGW-2 \cite{KGW2023} and multi-bit CTWL-5/CTWL-10 \cite{CTWL2024}. In KGW-1/KGW-2, the number is that the hash is determined by the first 1/2 tokens. In CTWL-5/CTWL-10, the number is that 1-bit watermark is embedded in 5/10 tokens, and the proxy model is GPT2-Large \cite{GPT22019}. {They can compare the generation effects of ITSmark from different aspects, and these performances have been widely recognized.}

	\subsubsection{Hyperparameters} In the comparison study, the Top\_k in the sampling is 40, and the Beam in the search is 4. Weight parameter $\lambda  = 1.0$, read the watermark at one time $\varepsilon  = 16$, the maximum generated text length is 1,024. {The encryption algorithm can be arbitrary, such as an encryption algorithm with high-performance key management. Here we use the simple and intuitive RSA encryption algorithm to demonstrate.} {In the ablation study, we calculate the “entropy” of each sentence, and then a part of the sentences is selected with large entropy to regenerate and embed the watermark in partial embedding supported by ITSmark.} “Partial embedding” ratio $\eta$ is set \{0.9, 0.7, 0.5, 0.3, 0.1\}, “Full embedding” ratio $\eta$ is 1.0, weight parameter $\lambda$ is set \{1.5, 1.2, 1.0, 0.9, 0.8, 0.7, 0.6, 0.5, 0.4\}, and read the watermark $\varepsilon$ is set \{20, 16, 12, 8\}.

	\setcounter{table}{3}
	\subsubsection{Experimental overview} We provide an overview of each part of the experiments, as shown in TABLE \ref{tab2}.

	\begin{table*}[htbp]
		\centering
		{\fontsize{7.4}{8}\selectfont
			\setlength{\tabcolsep}{0.75mm}
			\caption{Comparison of ITSmark and baselines regarding Perplexity (PPL), BERTScore (Score), and ROUGE. “Sample” and “Search” are un-watermarked data. “$\uparrow$” and “$\downarrow$” are the higher/lower the corresponding values, the better the performance. “\textbf{Bold}” is the best result, and “\underline{  *}” is the suboptimal result.}\label{tabb1}%
			\begin{tabular}{c||ccc|ccc|ccc|ccc||ccc}
				\toprule[1.2pt]
				\multicolumn{1}{c||}{\multirow{2}[2]{*}{\textbf{LLaMA2-7B}}} &\multicolumn{3}{c|}{TV \cite{KDD2023}} &\multicolumn{3}{c|}{BDD \cite{KDD2023}} & \multicolumn{3}{c|}{AlpacaFarm \cite{INT2023}} &  \multicolumn{3}{c||}{FinQA \cite{FinQA2018}} & \multicolumn{3}{c}{\textbf{Avg.} (LLMs)} \\
				\cmidrule{2-16}          & PPL $\downarrow$  & Score $\uparrow$ & ROUGE-L $\uparrow$ &PPL $\downarrow$  & Score $\uparrow$ & ROUGE-L $\uparrow$ &PPL $\downarrow$  & Score $\uparrow$ & ROUGE-L $\uparrow$ & PPL $\downarrow$  & Score $\uparrow$ & ROUGE-2 $\uparrow$ & PPL $\downarrow$  & Score $\uparrow$ & ROUGE $\uparrow$ \\
				\midrule
				\rowcolor[rgb]{ .900,  .900,  .900} Sample &5.52 &100.00&100.00&4.26 &100.00&100.00& 3.23  & 100.00 & 100.00  & 3.19   & 100.00  & 100.00  &4.05	&100.00&	100.00
				\\
				+KGW-1 &6.79 &67.30&40.89 &5.93 &69.12 &40.36 & 4.57  & 63.83  & 34.06  & 3.89  & 63.17  & 13.36  &5.30	&65.86&	32.17
				\\
				+KGW-2 &\underline{6.67*} &67.58&41.08 &\underline{5.89*} &\underline{69.58*} &41.17 & 4.45   & 64.01  & 34.91  & \underline{3.84*}  & 63.49  & 14.37  & \underline{5.21*}	&66.17&	32.88
				\\
				+CTWL-5 &7.61&68.91&45.65&6.42 &67.17 &40.22 & 4.51  & 64.93  & 42.19  & 4.36  & 65.96  & \underline{23.04*}  & 5.73&	66.74&	37.78
				\\
				+CTWL-10 &7.33 &\underline{69.39*}&\underline{46.34*}&6.14&68.69 &\underline{42.04*}& \underline{4.25*}  & \underline{66.17*}  & \underline{42.76*}  & 4.29  & \underline{66.60*}  & 22.99  & 5.50&	\underline{67.71*}&	\underline{38.53*}
				\\
				\textbf{+Ours} &\textbf{5.91} &\textbf{80.23} &\textbf{61.31} &\textbf{4.27} &\textbf{73.35} &\textbf{45.11} & \textbf{3.32}  & \textbf{76.07}  & \textbf{52.79}  & \textbf{3.40}  & \textbf{74.05}  & \textbf{31.78}  & \textbf{4.23}&	\textbf{75.93}&	\textbf{47.75}
				\\
				\midrule
				
				\rowcolor[rgb]{ .900,  .900,  .900} Search &5.43 &100.00&100.00&3.55 &100.00&100.00& 2.80  & 100.00  & 100.00  & 3.04  & 100.00  & 100.00  &3.71&	100.00&	100.00
				\\
				+KGW-1 &7.13 &69.81 &45.47 &5.06 &68.57 &40.88 & 4.47  & 65.16  & 34.99  & 4.61  & 63.39  & 14.31  & 5.32&	66.73&	33.91
				\\
				+KGW-2 &6.66 &70.02 &46.89 &4.92 &\underline{69.59*} &41.07 & 4.32  & 65.45  & 35.75  & 4.22  & 63.45  & 14.26  & 5.03&	67.13&	34.49
				\\
				+CTWL-5 &6.81&71.26&48.37&4.48&67.16&40.74 & 3.87  & 70.73  & 47.79  & 4.10  & 65.70  & 24.35  & 4.82	&68.71&	40.31
				\\
				+CTWL-10 &\underline{6.40*}&\underline{73.97*} &\underline{51.49*} &\underline{4.41*} &68.88 &\underline{41.29*} & \underline{3.72*}  & \underline{72.35*}  & \underline{49.15*} & \underline{3.93*}  & \underline{66.32*}  & \underline{26.36*}  &\underline{4.62*}&	\underline{70.38*}&	\underline{42.07*}
				\\
				\textbf{+Ours} &\textbf{5.90} &\textbf{80.50} &\textbf{62.53} &\textbf{4.18} &\textbf{72.09} &\textbf{43.99} & \textbf{3.20}  & \textbf{78.38}  & \textbf{57.09}  & \textbf{3.49}  & \textbf{73.97}  & \textbf{32.07}  &\textbf{4.19}&	\textbf{76.24}&	\textbf{48.92}
				\\
				\midrule[1.2pt]
				
				\multirow{2}[2]{*}{\textbf{ChatGLM3-6B}} &\multicolumn{3}{c|}{TV \cite{KDD2023}} &\multicolumn{3}{c|}{BDD \cite{KDD2023}}& \multicolumn{3}{c|}{AlpacaFarm \cite{INT2023}} & \multicolumn{3}{c||}{FinQA \cite{FinQA2018}} & \multicolumn{3}{c}{\textbf{Avg.} (LLMs)} \\
				\cmidrule{2-16}          & PPL $\downarrow$  & Score $\uparrow$ & ROUGE-L $\uparrow$ &PPL $\downarrow$  & Score $\uparrow$ & ROUGE-L $\uparrow$ &PPL $\downarrow$  & Score $\uparrow$ & ROUGE-L $\uparrow$ & PPL $\downarrow$  & Score $\uparrow$ & ROUGE-2 $\uparrow$ & PPL $\downarrow$  & Score $\uparrow$ & ROUGE $\uparrow$  \\
				\midrule
				\rowcolor[rgb]{ .900,  .900,  .900} Sample &6.61 &100.00&100.00&5.68&100.00&100.00& 4.11  & 100.00  & 100.00  & 4.22  & 100.00  & 100.00  & 5.16&	100.00&	100.00
				\\
				+KGW-1 &8.46 &68.47 &41.86 &8.17 &69.45   &38.33 & 5.85  & 63.54  & 34.98  & 6.08  & 61.98  & 14.83  & 7.14	&65.86&	32.50
				\\
				+KGW-2 &8.15 &69.53 &45.66 &8.01&70.15  &\underline{42.31*} & 5.70  & 63.91  & 34.12  & 5.30  & 62.90  & 15.62  & 6.79&	66.62&	34.43
				\\
				+CTWL-5 &7.39&74.08 &52.74 &7.73&69.10 &40.86 & 5.02  & 69.83  & 41.99  & 5.26  & 67.61  & 22.69  & 6.35&	70.16&	39.57
				\\
				+CTWL-10 &\underline{7.03*}&\underline{75.33*} &\underline{54.25*} &\underline{7.50*}&\underline{70.34*} &41.47& \underline{4.94*}  & \underline{70.18*}  & \underline{42.24*}  & \underline{5.14*}  & \underline{67.77*}  & \underline{22.74*}  & \underline{6.15*}&	\underline{70.91*}&	\underline{40.18*}
				\\
				\textbf{+Ours} &\textbf{6.96}&\textbf{78.74} &\textbf{60.32}&\textbf{6.31}&\textbf{71.68} &\textbf{45.17} & \textbf{4.36}  & \textbf{76.04}  & \textbf{52.63}  & \textbf{4.84}  & \textbf{73.97}  & \textbf{33.21}  &\textbf{5.62}&	\textbf{75.11}&	\textbf{47.83}
				\\
				\midrule
				
				\rowcolor[rgb]{ .900,  .900,  .900} Search &6.17&100.00&100.00&5.17&100.00&100.00& 3.83  & 100.00  & 100.00  & 4.38  & 100.00  & 100.00  & 4.89& 	100.00& 	100.00
				\\
				+KGW-1 &8.08 &66.39&40.96 &7.92&63.38&35.74 & 5.90  & 61.91  & 32.91  & 5.27  & 62.93  & 14.62  & 6.79&	63.65&	31.06
				\\
				+KGW-2 &7.73 &66.47&42.49 &7.76&63.45&36.18 & 5.67  & 62.09  & 33.06 & 5.12  & 63.02  & 15.96  &6.57&	63.76&	31.92
				\\
				+CTWL-5 &7.31&67.61&48.32&6.78&\underline{64.68*} &38.76 & 5.05  & 73.51  & 45.12  & 4.78  & 69.23  & 23.45  & 5.98&	68.76&	38.91
				\\
				+CTWL-10 &\underline{7.05*} &\underline{72.53*} &\underline{51.46*} &\underline{6.51*}&64.23 &\underline{40.79*} & \underline{4.89*}  & \underline{73.99*}  & \underline{45.26*}  & \underline{4.75*}  & \underline{69.37*}  & \underline{23.51*}  & \underline{5.80*}	&\underline{70.03*}&	\underline{40.26*}
				\\
				\textbf{+Ours} &\textbf{6.65}&\textbf{76.33} &\textbf{56.50} &\textbf{5.99}&\textbf{71.62} &\textbf{45.45} & \textbf{4.39}  & \textbf{77.39}  & \textbf{52.52}  & \textbf{4.68}  & \textbf{74.69}  & \textbf{33.11}  &\textbf{5.43}&	\textbf{75.01}&	\textbf{46.90}
				\\
				\midrule[1.2pt]
				
				\multirow{2}[2]{*}{\textbf{Avg.} (Datasets)} &\multicolumn{3}{c|}{TV \cite{KDD2023}} &\multicolumn{3}{c|}{BDD \cite{KDD2023}}& \multicolumn{3}{c|}{AlpacaFarm \cite{INT2023}} & \multicolumn{3}{c||}{FinQA \cite{FinQA2018}} & \multicolumn{3}{c}{\textbf{Avg.}} \\
				\cmidrule{2-16}          & PPL $\downarrow$  & Score $\uparrow$ & ROUGE-L $\uparrow$ &PPL $\downarrow$  & Score $\uparrow$ & ROUGE-L $\uparrow$ &PPL $\downarrow$  & Score $\uparrow$ & ROUGE-L $\uparrow$ & PPL $\downarrow$  & Score $\uparrow$ & ROUGE-2 $\uparrow$ & PPL $\downarrow$  & Score $\uparrow$ & ROUGE $\uparrow$  \\
				\midrule
				\rowcolor[rgb]{ .900,  .900,  .900} {\scriptsize Un-watermarked} &5.93&	100.00&	100.00&	4.67&	100.00&	100.00&	3.49&	100.00	&100.00&	3.71&	100.00&	100.00&	4.45&	100.00&	100.00
				\\
				+KGW-1 &7.62&	67.99&	42.30&	6.77&	67.63&	37.32	&5.20&	63.61&	34.24&	4.96&	62.87&	14.28&	6.14&	65.53&	32.41
				
				\\
				+KGW-2 &7.30&	68.40&	44.03&	6.65&	\underline{68.19}&	40.18&	5.04&	63.87&	34.46&	4.62&	63.22&	15.05&	5.90&	65.92&	33.43
				\\
				+CTWL-5 &7.28&	70.47&	48.77&	6.35&	67.03&	40.15&	4.61&	69.75&	44.27&	4.63&	67.13&	23.38&	5.72&	68.59&	39.14
				\\
				+CTWL-10 &\underline{6.95*}&	\underline{72.81*}	&\underline{50.89*}&	\underline{6.14*}&	68.04&	\underline{41.40*}&	\underline{4.45*}&	\underline{70.67*}&	\underline{44.85*}&	\underline{4.53*}&	\underline{67.52*}&	\underline{23.90*}&	\underline{5.52*}&	\underline{69.76*}&	\underline{40.26*}
				\\
				\textbf{+Ours} &\textbf{6.36}&	\textbf{78.95}&	\textbf{60.17}&	\textbf{5.19}&	\textbf{72.19}&	\textbf{44.93}&	\textbf{3.82}&	\textbf{76.97}&	\textbf{53.76}&	\textbf{4.10}&	\textbf{74.17}&	\textbf{32.54}&	\textbf{4.87}&	\textbf{75.57}&	\textbf{47.85}
				\\
				\bottomrule[1.2pt]
			\end{tabular}%
		}
	\end{table*}%

	\subsection{Comparison with Baselines on Quality}\label{sec521}
	In terms of quality, we use the widely used Perplexity, BERTScore, and ROUGE metrics \cite{KGW2023, CTWL2024, WaterBench2024, WM_survey2024}. We use the advanced LLaMA3-8B \cite{LLaMA32024} to calculate the Perplexity. LLaMA3-8B is the best performance representative among the currently available LLMs, so it is used here to calculate. The formula of Perplexity is: 
	
	\begin{equation}
	\small
	Perplexity = \exp \left( { - \frac{1}{N}\sum\limits_{i = 1}^N {\log P({x_i}|{x_1}, \cdots ,{x_{i - 1}})} } \right).
	\end{equation}
	
\begin{figure}[!htbp]
		\centering
		\includegraphics[width=0.99\linewidth]{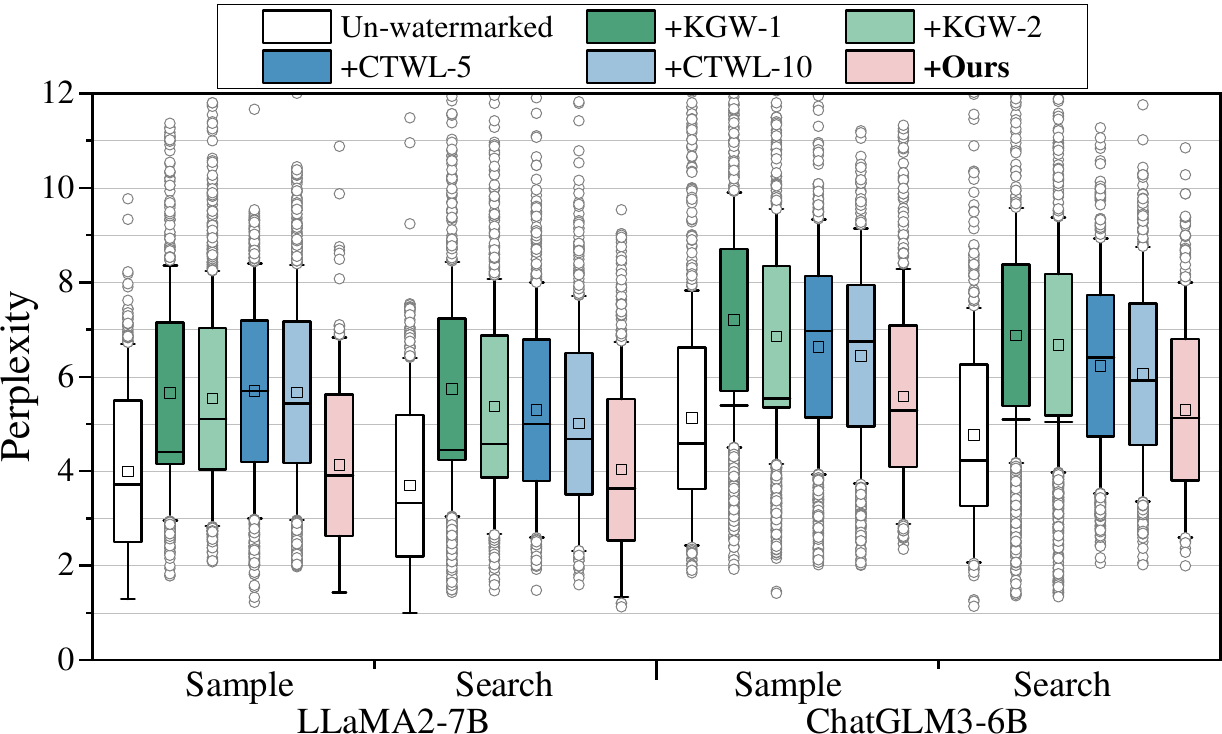}
		\caption{{Comparison of ITSmark (Ours) and baselines regarding Perplexity. The lower the Perplexity value, the better. The squares in the box represent the mean value, the horizontal line in the box represents the median value, and the gray circle represents the outlier.}}\label{fig4}
	\end{figure}

\begin{figure}[!htbp]
		\centering
		\includegraphics[width=0.98\linewidth]{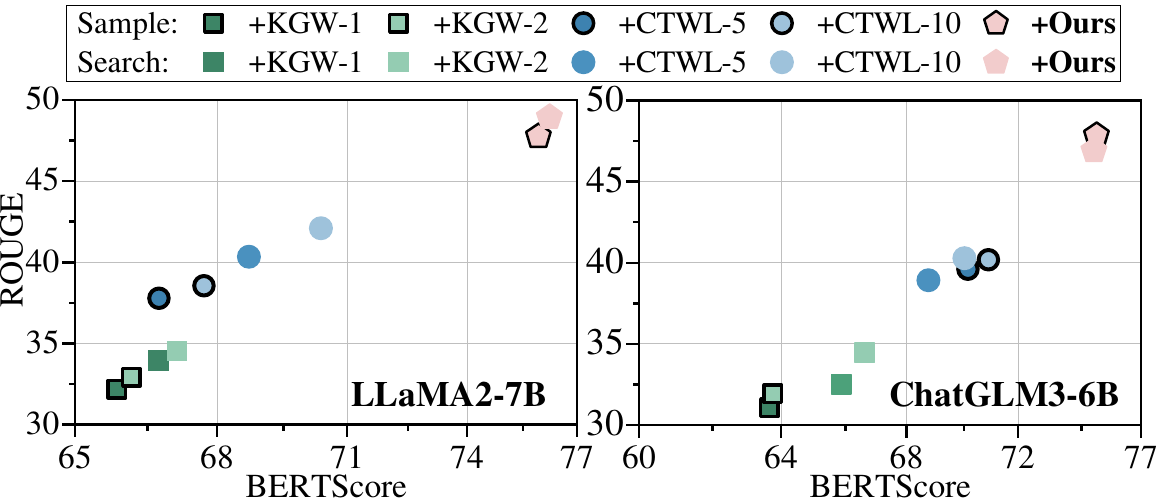}
		\caption{{Comparison of ITSmark (Ours) and baselines regarding BERTScore and ROUGE. The higher the BERTScore and ROUGE values, the better.}}\label{fig5}
	\end{figure}

	\noindent The lower the $Perplexity$, the quality of the generated text is higher. BERTScore \cite{BERTScore2020} considers the semantic similarity between the watermarked text and the text without the watermark. ROUGE \cite{ROUGE2004} focuses on the precision and acceptability of the watermarked text. Specifically, in the TV, BDD, and AlpacaFarm datasets, ROUGE-L is used to evaluate coherence and completeness; in the FinQA dataset, ROUGE-2 is used to evaluate the precision of text details. The higher the BERTScore and ROUGE, the quality of the generated text is higher. Fig. \ref{fig4} and Fig. \ref{fig5} show the comparison of ITSmark and baselines on Perplexity, BERTScore, and ROUGE. For detailed results on the three datasets, please see TABLE \ref{tabb1}.

	It can be seen from Fig. \ref{fig4} and Fig. \ref{fig5} that compared with the advanced baselines, ITSmark has lower Perplexity, higher BERTScore, and higher ROUGE. It indicates that the watermarked text generated by ITSmark has higher quality.
	
	\subsection{Comparison with Baselines on Embedding Time and Extraction Success Rate}\label{sec522}
	Watermarking embeds a watermark with practical meaning, so the focus of watermarking is the accuracy of extraction. Only the entire extraction can be called successful extraction. We are expected to not consume too much time to obtain watermarked text and to be able to protect the integrity of copyright.
	
	Since the KGW scheme is a zero-bit watermarking, it is impossible to embed and extract specific watermark information. Therefore, the experiments in this regard compare the ITSmark and CTWL multi-bit watermark schemes. ITSmark and CTWL, which can embed custom information, are compared in terms of embedding time and extraction success rate. The results are shown in TABLE \ref{tab3}.

	\begin{table}[!htbp]
		\centering
		\setlength{\tabcolsep}{1.1mm}
		\caption{Comparison of ESMark (Ours) and baselines regarding embedding time and success rate. The extraction and embedding time of ESMark are almost the same. “\textbf{Bold}” is the best result, and “\underline{  *}” is the suboptimal result.}\label{tab3}%
		\begin{tabular}{c||cccc||c}
			\toprule[1.2pt]
			\multicolumn{1}{c||}{Embedding time (sec) $\downarrow$} & TV&BDD&AlpacaFarm &  \multicolumn{1}{c||}{FinQA} & \textbf{Avg.} \\
			\midrule
			+CTWL-5 &2.02&2.90& 3.69 & 3.78 & 3.10
			\\
			+CTWL-10 &2.15&2.94& 3.41  & 3.04 & \underline{2.89*}
			\\
			\textbf{+Ours}  &1.47&2.68& 3.16 & 2.87 & \textbf{2.55} \\
			\midrule[1.2pt]
			Success rate (\%) $\uparrow$ &TV&BDD& AlpacaFarm & FinQA & \textbf{Avg.} \\
			\midrule
			+CTWL-5 &79.85&73.43& 71.16 &84.50  & 77.24
			\\
			+CTWL-10 &91.16&93.68& 92.38 &97.52 & \underline{93.69*}
			\\
			\textbf{+Ours}  &100.00&100.00& 100.00   & 100.00   & \textbf{100.00} \\
			\bottomrule[1.2pt]
		\end{tabular}%
	\end{table}%

	From TABLE \ref{tab3}, we can find that in different datasets, ITSmark takes a shorter average time to generate watermarked data and ensure the entire extraction. It shows that ITSmark can protect the integrity of copyright and improve the reliability of copyright verification. {The CTWL scheme cannot guarantee the complete extraction of watermark information because it introduces an additional watermark proxy model in the extraction process, and its embedding and extraction are not completely inverse, so complete extraction cannot be guaranteed.} {{In addition, the complete extraction of ITSmark is not affected by the key parameters $\lambda$, $\varepsilon$, and $\eta$ of the scheme, because its embedding and extraction processes are inverse and independent of the parameters.}}

	\subsection{Comparison with Baselines on Unforgeability}\label{sec523}
\begin{figure}[!htbp]
		\centering
		\includegraphics[width=0.99\linewidth]{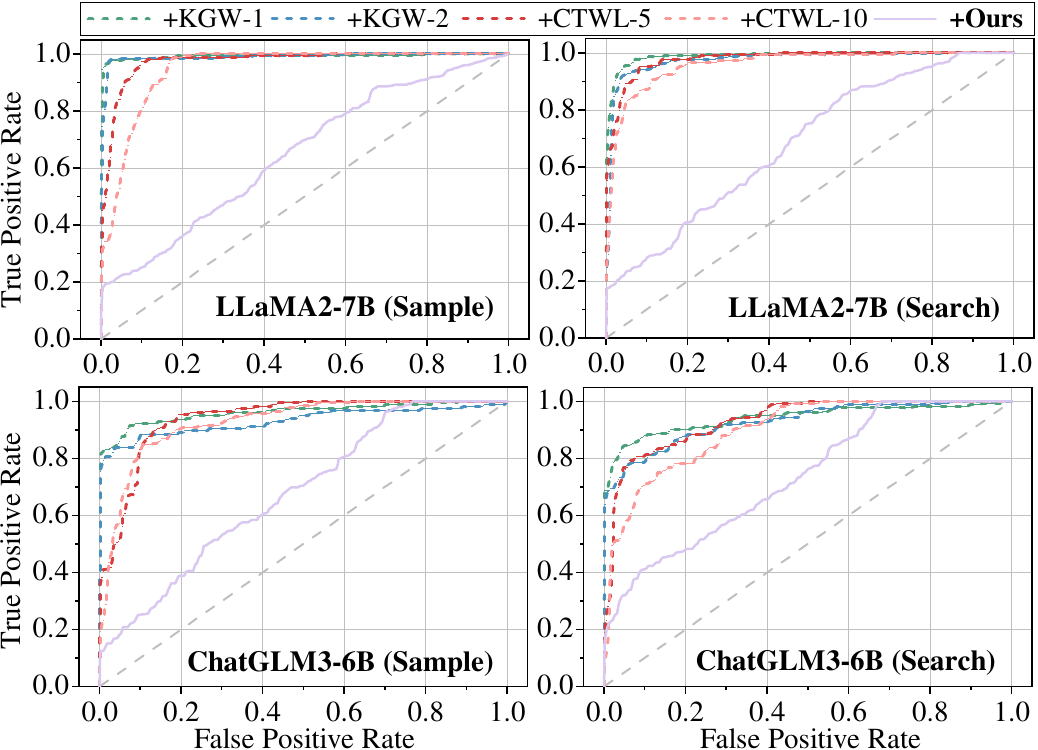}
		\caption{Comparison of ITSmark and baselines regarding the unforgeability in private detection. The smaller the AUC, the stronger the unforgeability.}\label{fig6}
	\end{figure}

	As Liu et al. \cite{WM_survey2024} mentioned, in the real world, once illegal persons decipher the way of generating watermarked data, they can easily forge or delete the existing watermark. This greatly jeopardizes the reliability of copyright verification. Since the extraction way of ITSmark is only accessible to specific institutions, it is necessary to ensure the unforgeability in private detection. This unforgeability is largely derived from the imperceptibility of the watermark, that is, the illegal persons do not know whether the data contains a watermark. We use part of the un-watermarked and watermarked data to train the RoBERTa \cite{RoBERTa2019}, and the rest for testing. The test results are shown in the ROC-AUC. The comparison with baselines on the unforgeability is shown in Fig. \ref{fig6}.

	The results in Fig. \ref{fig6} show that compared with baselines, ITSmark is difficult to detect by the classifier, which improves the imperceptibility and thus ensures the unforgeability. {Furthermore, this result also indirectly shows that the watermarked data obtained by ITSmark is more like the original un-watermarked data. This is because RoBERTa can capture the statistical distribution features of watermarked and un-watermarked data. The ROC-AUC curve of watermarked data obtained by ITSmark is closer to the $x=y$ dashed line than that of the baseline schemes, so it is more difficult to be distinguished by the RoBERTa classifier, that is, the watermarked data obtained has a statistical distribution that is more consistent with the original un-watermarked data.}

	\subsection{Permission verification failed}\label{sec531}
\begin{figure}[!htbp]
		\centering
		\includegraphics[width=0.99\linewidth]{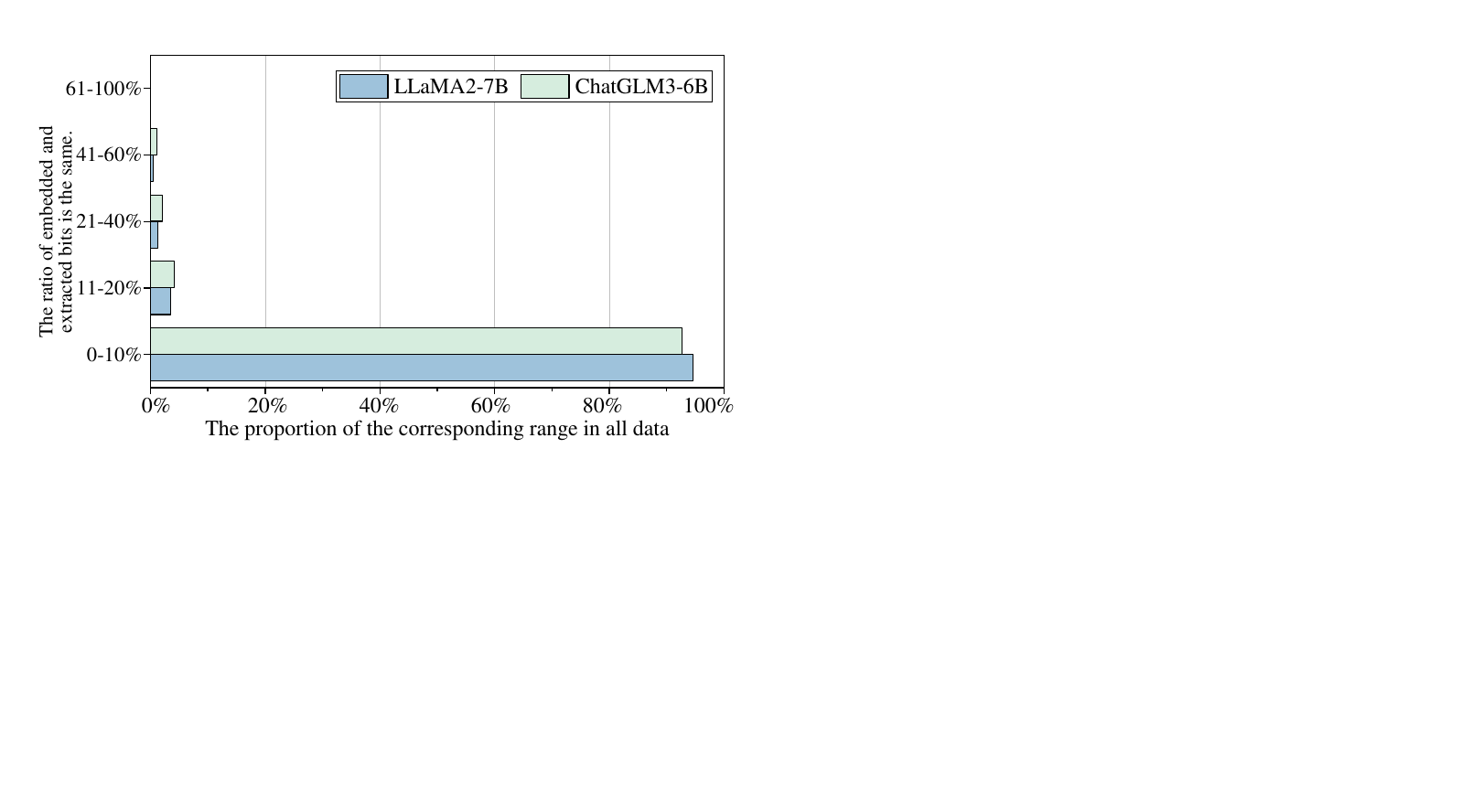}
		\caption{{{The extraction results when the permission verification fails. The vertical axis represents the ratio of extracted and embedded watermark information, which is the same as that of watermarked text. The horizontal axis represents the proportion of text items within a certain range to the total number of items.} For example, if the embedded 9-bit is “010110010” and “0” is extracted, “The ratio of embedded and extracted bits is the same” is 11.11\%. “The ratio of embedded and extracted bits is the same.” of 0-10\% exceeds 90\% of all data.}}\label{fig7}
	\end{figure}

	To ensure the security and authorization of watermark extraction, ITSmark needs to perform permission verification before extraction. That is, only the watermark data and the extraction way cannot entirely restore the watermark. The extractor needs the correct cipher data and private key provided by the sender to successfully pass the permission verification, so that the watermark can be entirely extracted. Fig. \ref{fig7} shows the extraction results when the permission verification fails.

	\begin{table*}[!htbp]
		\centering
		{\fontsize{7}{8}\selectfont
			\setlength{\tabcolsep}{0.7mm}
			\setlength{\fboxsep}{0pt}
			\caption{{Example of watermark extraction when permission verification is passed and failed. After decrypting the ciphertext data, “P” represents the Prompt field, and “R” represents the number of bits to be read at one-time field.}}\label{tabe1}%
			\begin{tabular}{p{18.8em}||p{33em}|p{5em}|p{8em}|p{4.5em}}
				\toprule[1.2pt]
				\multicolumn{1}{c||}{Prompt} & \multicolumn{1}{c|}{Watermarked Data} & \multicolumn{1}{c|}{Cipher Data} & \multicolumn{1}{c|}{Decrypting Cipher} & \multicolumn{1}{c}{Extraction}\\
				\midrule[0.5pt]
				The POV car is swerving wildly back and forth on a two lane road. The POV car makes a large turn to the left and proceeds to drive backwards briefly in the opposite lane. The POV car then swerves a bit further causes it to fall into a ditch on the side of the road. The POV car continues moving fairly quickly in the ditch before slowly reducing the speed. Which could be the reason for this accident?&
				The reason for the accident could be that the driver of the POV car was distracted or impaired while driving, causing them to lose control of the vehicle and swerve wildly. The sudden turn to the left and drive backwards in the opposite lane may have been an attempt to correct the vehicle's position, but it ultimately led to the car falling into a ditch on the side of the road. The fact that the car continued moving fairly quickly in the ditch suggests that the driver may have been unable to regain control of the vehicle or may have been in a state of disorientation after the accident.
				
				\textbf{(Embedded watermark: 1011100)}
				&
				\begin{verbatim}\xb7Z\xb\end{verbatim}
				\begin{verbatim}f\xc6\xc\end{verbatim}
				\begin{verbatim}2\xbc\xd\end{verbatim}
				\begin{verbatim}2@s\x...\end{verbatim}
				&
				\multicolumn{1}{p{8em}|}{\multirow{1}{=}{ \newline{}\newline{} 'P': The POV car \newline{} is swerving wildly \newline{}back and forth on a \newline{} two lane road. ...\newline{} 'R': 16; '$\lambda$': 1.0}}
				&
				\multicolumn{1}{p{4.5em}}{\multirow{1}{=}{ \newline{}\newline{}\newline{} 1011100 \newline{}(Accuracy: \newline{}100\%)}}
				\\
				\midrule[0.5pt]
				\multicolumn{1}{p{18em}||}{\multirow{1}{=}{ \newline{}\newline{} What are the names of some famous actors \newline{} that started their careers on Broadway?}}
				&Some famous actors who started their careers on Broadway include:
				
				* Hugh Jackman
				* Nicole Kidman
				* Daniel Radcliffe
				* Idina Menzel
				* James Corden
				* Lin-Manuel Miranda
				* John Leguizamo
				* Sarah Jessica Parker
				* Bernadette Peters
				
				These actors have all made a name for themselves on the Great White Way before going on to successful careers in film and television.
				
				\textbf{(Embedded watermark: 100100001)}
				&\begin{verbatim}\x88\x1b\end{verbatim}
				\begin{verbatim}\x07\...\end{verbatim}
				&'P': What are the names of some famous actors that started their careers on Broadway?
				
				'R': 16; '$\lambda$': 1.0
				&
				\multicolumn{1}{p{4.5em}}{\multirow{1}{=}{ \newline{}\newline{} 100100001 \newline{}(Accuracy: \newline{}100\%)}}
				\\
				\midrule[0.5pt]
				\multicolumn{1}{p{18em}||}{\multirow{1}{=}{ \newline{}\newline{}\newline{}\newline{}\newline{}\newline{}\newline{} Traffic begins to move, what will the car do \newline{} next? Please give me a detailed plan.}}
				&
				As traffic begins to move, the car will first check its surroundings to ensure that it has enough space to merge into the flow of traffic. Once it has confirmed that there is enough space, the car will slow down to a safe speed and begin to merge into the flow of traffic.
				
				Next, the car will use its turn signal to indicate its intention to merge and will check for any oncoming traffic or pedestrians in the area. Once it has confirmed that it has a clear path, the car will make a smooth and gradual merge into the flow of traffic.
				
				Once the car has merged into the flow of traffic, it will adjust its speed to match the flow of traffic around it and maintain a safe distance from the vehicle in front of it. The car will also continue to monitor its surroundings and be prepared to adjust its speed and position as needed to maintain safety on the road.
				
				Overall, the car will take a cautious and deliberate approach to merging into traffic, prioritizing safety above all else.
				
				\textbf{(Embedded watermark: 10110100001)}
				&\multicolumn{1}{p{5.3em}|}{\multirow{1}{=}{ \newline{}\newline{}\newline{}\newline{}\newline{}\newline{}\newline{} If cipher data \newline{} is wrong.}}
				&\multicolumn{1}{p{8em}|}{\multirow{1}{=}{ \newline{}\newline{}\newline{}\newline{}\newline{}\newline{}\newline{}\newline{} Wrong}}&
				\multicolumn{1}{p{4.5em}}{\multirow{1}{=}{ \newline{}\newline{}\newline{}\newline{}\newline{}\newline{}\newline{} 0 \newline{}(Accuracy: \newline{}0\%)}}
				\\
				\bottomrule[1.2pt]
			\end{tabular}%
		}
	\end{table*}%

	The results in Fig. \ref{fig7} show that if the verification fails, the watermark cannot be entirely extracted, which ensures the security and authorization of the extraction. {In addition, we also give some examples of extraction when the verification succeeds and fails. The results are shown in TABLE \ref{tabe1}.}

	\subsection{Traceability of tampered locations}\label{sec532}
	If the verification is passed, and the watermark still cannot be entirely extracted, it means that the content has been tampered with. {ITSmark can effectively track and locate the tampered location without knowing the original watermarked data. That is, the receiver can trace the tampered location by only receiving the tampered watermarked data.
		
		Since a word list to be selected, that is, a candidate pool, will be constructed during the process of embedding the watermark, the size of this candidate pool is set to Top\_k (ITSmark sets it to 40) in this paper, that is, no matter what the watermark information is, the determined token must be one of these Top\_k. When the watermarked data is tampered with (using RoBERTa replacement or GPT4 rewriting), the conditional probability of these tampered tokens may not be in the Top\_k, and these tokens must have been tampered with. In addition, even in the Top\_k, during the ITSmark design process, the interval bias will be assigned to several tokens with large conditional probabilities, so the probability of selecting these tokens is extremely high. That is, some tokens ranked lower in the Top\_k will hardly be selected by ITSmark, so if such a token appears, it is considered to be tampered with with a high probability.}
	
	We designed a simple discrimination method to trace tampered locations. The remaining identified tampering probabilities (TP) are as follows:
	
	\begin{equation}
	\small
	\text{TP}(x_i) =
	\begin{cases} 
	1, & o(x_i) > \text{Top\_k} \\
	0.75, & \text{Top\_k}/2 < o(x_i) \leq \text{Top\_k} \\
	0.5, & \text{Top\_k}/4 < o(x_i) \leq \text{Top\_k}/2 \\
	0.3, & \text{Top\_k}/8 < o(x_i) \leq \text{Top\_k}/4 \\
	0, & o(x_i) \leq \text{Top\_k}/8
	\end{cases}.
	\end{equation}

	\begin{table*}[htbp]
		\centering
		\setlength{\tabcolsep}{1mm}
		\setlength{\fboxsep}{0pt}
		\caption{Examples of tampering and traceability. The “Instruction” is derived from AlpacaFarm and FinQA datasets. The “Watermarked Data” is obtained using the ITSmark full embedding. \textcolor{red}{Red} represents locations that have been tampered with. \colorbox[RGB]{211,240,218}{The background} represents locations that are considered tampered with by the above strategy. The darker the background color, the higher the probability that it has been tampered with.}\label{taba}%
		\begin{tabular}{p{31em}|p{31.7em}}
			\toprule[1.2pt]
			\multicolumn{1}{c|}{Watermarked Data} & \multicolumn{1}{c}{\textcolor{red}{\textbf{Attack}} and \colorbox[RGB]{211,240,218}{Tracing}} \\
			\midrule[0.5pt]
			The first smartphones were created in the late 1990s and early 2000s. The first commercial smartphone was the Palm handheld device, which was released in 1999. & The first smartphones were created in the late 1990s and early 2000s. The first \textcolor{red}{\colorbox[RGB]{86,196,110}{\textbf{mainstream}}} smartphone was the Palm handheld device, which was released in 1999.\\
			\midrule[0.5pt]
			Interesting! 'The Twelve Caesars' by Suetonius was written in the late 1st century AD, specifically between 75 and 80 AD. The book is a collection of biographies of the Roman emperors from Julius Caesar to Domitian, and it provides valuable insights into the political and social climate of ancient Rome during that time. Suetonius was a Roman historian and politician who lived during the 1st and 2nd centuries AD, and his work is considered one of the most important sources of information about the Roman Empire during this period. & \textcolor{red}{\colorbox[RGB]{186,231,196}{\textbf{Interest}}} \textcolor{red}{\colorbox[RGB]{86,196,110}{\textbf{reading}}}! 'The \textcolor{red}{\colorbox[RGB]{86,196,110}{\textbf{Seven}}} \textcolor{red}{\colorbox[RGB]{86,196,110}{\textbf{Caetans}}}' by Suetonius was written in the \textcolor{red}{\textbf{early}} 1st century \textcolor{red}{\colorbox[RGB]{186,231,196}{\textbf{BC}}}, specifically between 75 and 80 AD. The book is \textcolor{red}{\colorbox[RGB]{186,231,196}{\textbf{the}}} \colorbox[RGB]{186,231,196}{collection} of biographies of twelve Roman emperors from Julius Caesar to \textcolor{red}{\colorbox[RGB]{86,196,110}{\textbf{Sumitian}}}, and \textcolor{red}{\colorbox[RGB]{86,196,110}{\textbf{thus}}} \textcolor{red}{\colorbox[RGB]{186,231,196}{\textbf{offers}}} valuable insights into the political and social climate of ancient Rome during that time. Suetonius was \textcolor{red}{\colorbox[RGB]{186,231,196}{\textbf{the}}} Roman historian \textcolor{red}{\colorbox[RGB]{86,196,110}{\textbf{or}}} politician who lived during the 1st and \textcolor{red}{\colorbox[RGB]{186,231,196}{\textbf{2rd}}} centuries AD, and his work is considered one \textcolor{red}{\colorbox[RGB]{110,204,131}{\textbf{among}}} \textcolor{red}{\colorbox[RGB]{110,204,131}{\textbf{our}}} \textcolor{red}{\colorbox[RGB]{110,204,131}{\textbf{more}}} important sources of \textcolor{red}{\colorbox[RGB]{186,231,196}{\textbf{knowledge}}} about the Roman Empire during this \colorbox[RGB]{186,231,196}{period\textcolor{red}{\textbf{!}}} \\
			\midrule[0.5pt]
			Walter Elias Disney, known professionally as Walt Disney, was born on December 5, 1901, in Chicago, Illinois. He was an American entrepreneur, animator, and filmmaker who is best known for his influence in the entertainment industry, particularly in the field of animation. & Walter Elias Disney, known professionally as Walt \textcolor{red}{\colorbox[RGB]{186,231,196}{\textbf{Mouse}}}, was born on December 5, 1901, \textcolor{red}{\colorbox[RGB]{186,231,196}{\textbf{near}}} Chicago, Illinois. He was an American entrepreneur, animator, and filmmaker who is best known for his influence in the entertainment industry, particularly in \textcolor{red}{\colorbox[RGB]{156,220,171}{\textbf{his}}} \colorbox[RGB]{156,220,171}{field} of animation. \\
			Disney grew up in a family of five children and developed an early interest in drawing and storytelling. He began his career as an animator in the early 1920s, working for several animation studios before eventually forming his own company, Disney Brothers Cartoon Studio, in 1923. & Disney grew up in a family of five children and developed \textcolor{red}{\textbf{his}} early interest in drawing and storytelling. \textcolor{red}{\colorbox[RGB]{86,196,110}{\textbf{Disney}}} began his career as an animator in the early 1920s \textcolor{red}{\colorbox[RGB]{156,220,171}{\textbf{while}}} working for several animation studios before eventually forming his own company, Disney Brothers Cartoon Studio, in 1923. \\
			Throughout his career, Disney was known for his innovative approach to animation, which included the development of new techniques and technologies. He is credited with creating some of the most beloved and enduring animated characters in history, including Mickey Mouse, Donald Duck, and Goofy. & Throughout his career, Disney was known for his innovative approach to animation, which included \textcolor{red}{\colorbox[RGB]{186,231,196}{\textbf{his}}} development of new techniques and technologies. He is credited with creating some of the most beloved and enduring animated characters in \textcolor{red}{\colorbox[RGB]{110,204,131}{\textbf{animation}}}, including Mickey Mouse, Donald Duck, and \textcolor{red}{\colorbox[RGB]{86,196,110}{\textbf{Doofy}}}. \\
			In addition to his work in animation, Disney was also a successful film producer and entrepreneur. He founded the Disneyland theme park in 1955 and expanded his company into television, film, and other ventures. & \colorbox[RGB]{186,231,196}{In} addition to his work in animation, Disney was also a successful film producer and entrepreneur. He founded the Disneyland theme park in 1955 and expanded his company into television \textcolor{red}{\textbf{and}} film, \colorbox[RGB]{186,231,196}{and} \colorbox[RGB]{186,231,196}{other} ventures. \\
			Disney passed away on December 15, 1966, but his legacy continues to be celebrated through the Disney Company, which remains a leader in the entertainment industry today. & Disney passed away on December 15, 1966, but his legacy continues to be celebrated through the Disney Company, which remains \textcolor{red}{\colorbox[RGB]{186,231,196}{\textbf{the}}} leader in the entertainment industry today. \\
			\bottomrule[1.2pt]
		\end{tabular}%
	\end{table*}%

	\noindent where, $o(x_i)$ represents the rank of the conditional probability of token $x_i$ in the current conditional probability distribution. This discrimination method is relatively efficient and concise, and the traceability of each text only takes less than 0.5 seconds. Please note that the strategy here is just a simple presentation, and a better tracing strategy is worth further exploration by scholars.

	\begin{figure}[!htbp]
		\centering
		\includegraphics[width=0.99\linewidth]{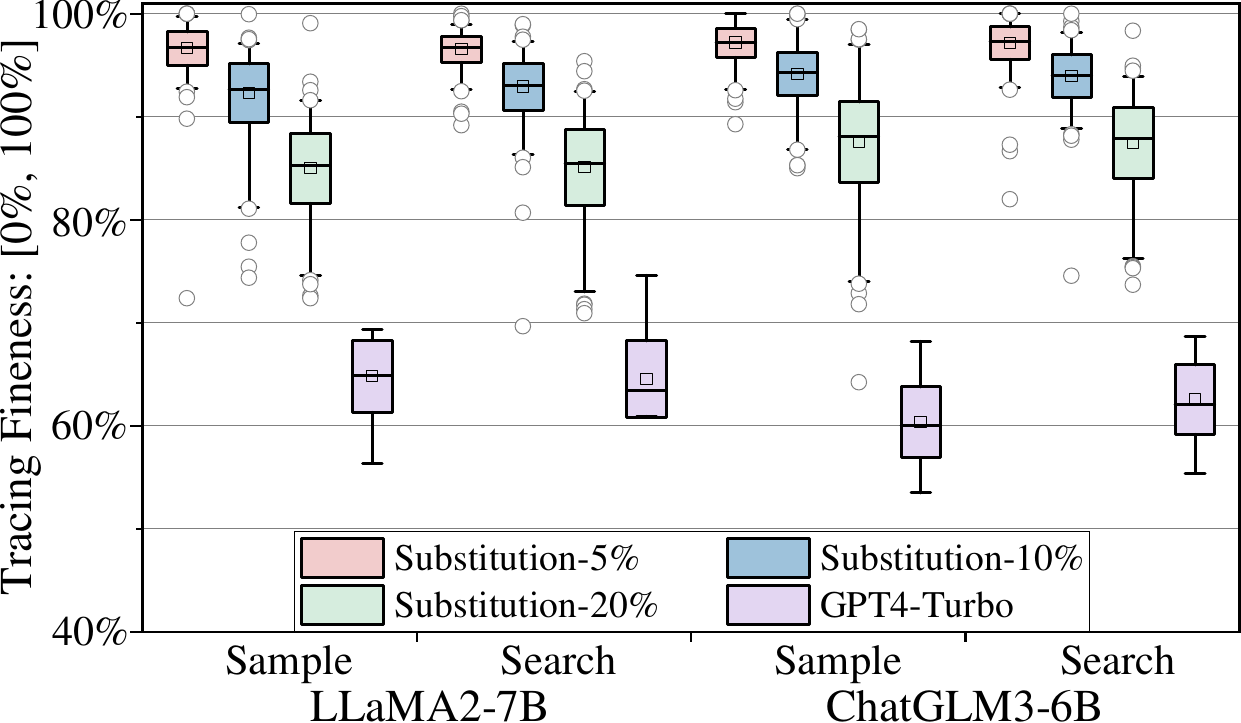}
		\caption{{Fineness of tampered location tracing. The value range is [0\%, 100\%].} The squares in the box represent the mean value, the horizontal line in the box represents the median value, and the gray circle represents the outlier. The fineness is calculated using the product of the probability and whether the real data has been tampered with. {For example, if the token conditional probability ranking of a certain position is 35th (i.e. Top\_k/2), ITSmark believes that it has a 75\% probability of being tampered, and it indeed tampers, that is, the label is 1. Then its fineness component score is $75\%\times 1=75\%$. Finally, the fineness component scores of all tokens are averaged to get the final “Tampering Fineness” value.}}\label{fig8}
	\end{figure}

	Here we use word-level substitution attacks and document-level rewrite attacks. The substitution attack uses RoBERTa \cite{RoBERTa2019} to replace (5\%, 10\%, and 20\%) tokens of the text while maintaining the integrity of the overall semantics. The rewrite attack uses GPT4-Turbo \cite{GPT42023} to rewrite the text with the instruction: “\textit{Please rewrite this text}:”. The performance and examples of tampered location tracing are shown in Fig. \ref{fig8} and TABLE \ref{taba}.
	
	From the results in Fig. \ref{fig8} and TABLE \ref{taba}, we can see that after the watermarked data obtained by ITSmark tampers, we can cleverly use the ITSmark embedding rules to design an algorithm for tracing the tampering location and accurately trace the tampering location. This provides a reliable basis for anti-tampering and tampering forensics.

	\subsection{Ablation on Embedding Ratios $\eta$}\label{sec541}
\begin{figure}[!htbp]
		\centering
		\includegraphics[width=0.99\linewidth]{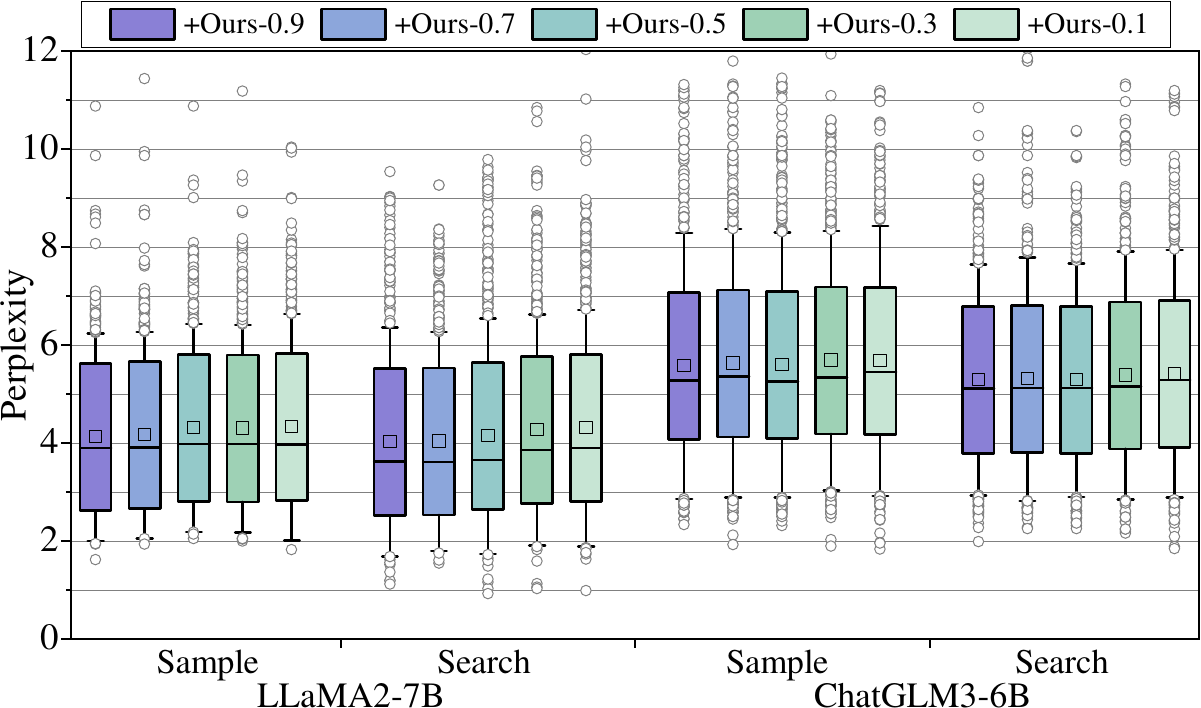}
		\caption{{Comparison of Perplexity under different embedding ratios $\eta$. The squares in the box represent the mean value, the horizontal line in the box represents the median value, and the gray circle represents the outlier.}}\label{fig9}
	\end{figure}

\begin{figure}[!htbp]
		\centering
		\includegraphics[width=0.99\linewidth]{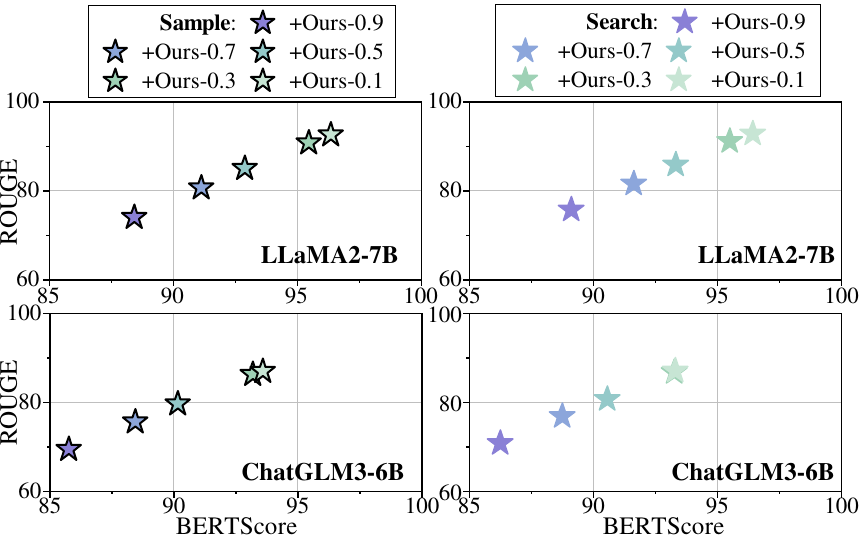}
		\caption{{Comparison of BERTScore and ROUGE under different $\eta$.}}\label{fig10}
	\end{figure}

	The embedding ratios $\eta  = \{0.9,0.7,0.5,0.3,0.1\}$, denoted as Ours-0.9, Ours-0.7, Ours-0.5, Ours-0.3, and Ours-0.1. Fig. \ref{fig9} and Fig. \ref{fig10} show the comparison of data quality under different $\eta$. The complete results are shown in TABLE \ref{tabb2}.

	\begin{table*}[htbp]
		\centering
		{\fontsize{7.4}{8}\selectfont
			\setlength{\tabcolsep}{0.7mm}
			\caption{Comparison of embedding ratios $\eta$ regarding Perplexity (PPL), BERTScore (Score), and ROUGE. “$\uparrow$” and “$\downarrow$” represent that the higher/lower the corresponding values, the better the performance. “\textbf{Bold}” is the best result, and “\underline{  *}” is the suboptimal result.}\label{tabb2}%
			\begin{tabular}{c||ccc|ccc|ccc|ccc||ccc}
				\toprule[1pt]
				\multirow{2}[2]{*}{\textbf{LLaMA2-7B}} &  \multicolumn{3}{c|}{TV \cite{KDD2023}} &\multicolumn{3}{c|}{BDD \cite{KDD2023}} & \multicolumn{3}{c|}{AlpacaFarm \cite{INT2023}} &  \multicolumn{3}{c||}{FinQA \cite{FinQA2018}} & \multicolumn{3}{c}{\textbf{Avg.} (LLMs)} \\
				\cmidrule{2-16}          &  PPL $\downarrow$  & Score $\uparrow$ & ROUGE-L $\uparrow$ &PPL $\downarrow$  & Score $\uparrow$ & ROUGE-L $\uparrow$ &PPL $\downarrow$  & Score $\uparrow$ & ROUGE-L $\uparrow$ & PPL $\downarrow$  & Score $\uparrow$ & ROUGE-2 $\uparrow$ & PPL $\downarrow$  & Score $\uparrow$ & ROUGE $\uparrow$  \\
				
				\midrule
				\rowcolor[rgb]{ .900,  .900,  .900} Sample &5.52 &100.00&100.00&4.26 &100.00&100.00& 3.23  & 100.00  & 100.00  &  3.19  & 100.00  & 100.00  & 4.05&	100.00&	100.00  \\
				+Ours-0.9 &5.88&90.48 &80.74 &4.28&92.42 &84.81 & 3.51  & 85.31  & 71.39  & 3.52  & 85.48  & 59.32  & 4.30&	88.42&	74.07 \\
				+Ours-0.7 &5.81&92.38 &84.64 &4.27&93.97 &87.89 & 3.42  & 88.82  & 79.34  &  3.51  & 89.38  & 70.98  & 4.25&	91.14&	80.71 \\
				+Ours-0.5 &5.81&94.15 &87.87 &\underline{4.26*}&94.69 &89.34 & 3.39  & 91.10  & 84.28  & 3.41  & 91.61  & 78.57  & 4.22&	92.89&	85.02  \\
				+Ours-0.3 &\underline{5.66*}&\underline{95.62*} &\underline{91.06*} &\underline{4.26*}&\underline{96.25*} &\underline{92.24*} & \underline{3.25*}  & \underline{94.59*}  & \underline{90.96*}  & \underline{3.30*}  & \underline{95.39*}  & \underline{88.82*}  & \underline{4.12*}&	\underline{95.46*}&	\underline{90.77*} \\
				+Ours-0.1 &\textbf{5.63} &\textbf{95.83} &\textbf{91.60} &\textbf{4.25}&\textbf{96.58} &\textbf{92.77} & \textbf{3.21}  & \textbf{96.07}  & \textbf{93.77}  & \textbf{3.23}  & \textbf{96.91}  & \textbf{92.21}  & \textbf{4.08}&	\textbf{96.35}	&\textbf{92.59}  \\	
				
				\midrule
				\rowcolor[rgb]{ .900,  .900,  .900} Search &5.43 &100.00&100.00&3.55 &100.00&100.00& 2.80  & 100.00  & 100.00  & 3.04  & 100.00  & 100.00  & 3.71&	100.00&	100.00
				\\
				+Ours-0.9 &5.87&92.40 &84.38 &4.18&92.31 &84.41 & 3.31  & 85.42  & 72.24 & 3.42  & 86.35  & 61.75  & 4.20&	89.12&	75.70
				\\
				+Ours-0.7 &5.80&94.38 &88.54 &4.07&93.62 &87.16 & 3.31  & 88.90  & 79.27 & 3.39  & 89.63  & 71.47  & 4.14	&91.63&	81.61
				\\
				+Ours-0.5 &5.64&94.80 &89.44 &3.95&94.92 &89.64 & 3.13  & 91.82  & 85.86 & 3.28  & 91.76  & 78.31  & 4.00	&93.33&	85.81
				\\
				+Ours-0.3 &\underline{5.52*}&\underline{96.27*} &\underline{92.40*} &\underline{3.93*}&\underline{96.04*} &\underline{92.04*} & \underline{3.08*}  & \underline{94.69*}  & \underline{91.82*} & \underline{3.20*}  & \underline{95.02*}  & \underline{88.16*}  & \underline{3.93*}&	\underline{95.51*}&	\underline{91.11*}
				\\
				+Ours-0.1 &\textbf{5.46} &\textbf{96.52} &\textbf{92.84} &\textbf{3.76}&\textbf{96.30} &\textbf{92.22} & \textbf{3.01}  & \textbf{96.23}  & \textbf{94.23} & \textbf{3.11}  & \textbf{96.64}  & \textbf{91.77}  & \textbf{3.84} &	\textbf{96.42} &	\textbf{92.77}
				\\	
				
				\midrule[1pt]
				\multirow{2}[2]{*}{\textbf{ChatGLM3-6B}} & \multicolumn{3}{c|}{TV \cite{KDD2023}} &\multicolumn{3}{c|}{BDD \cite{KDD2023}} & \multicolumn{3}{c|}{AlpacaFarm \cite{INT2023}} &  \multicolumn{3}{c||}{FinQA \cite{FinQA2018}} & \multicolumn{3}{c}{\textbf{Avg.} (LLMs)} \\
				\cmidrule{2-16}          &  PPL $\downarrow$  & Score $\uparrow$ & ROUGE-L $\uparrow$ &PPL $\downarrow$  & Score $\uparrow$ & ROUGE-L $\uparrow$ &PPL $\downarrow$  & Score $\uparrow$ & ROUGE-L $\uparrow$ & PPL $\downarrow$  & Score $\uparrow$ & ROUGE-2 $\uparrow$ & PPL $\downarrow$  & Score $\uparrow$ & ROUGE $\uparrow$  \\
				\midrule
				\rowcolor[rgb]{ .900,  .900,  .900} Sample &6.61 &100.00&100.00&5.68&100.00&100.00& 4.11  & 100.00  & 100.00  & 4.22  & 100.00  & 100.00  & 5.16&	100.00&	100.00
				\\
				+Ours-0.9 &6.92 &84.48 &69.22 &6.28 &84.36 &69.45 & 4.29  & 86.54  & 73.79 & 4.71  & 87.67  & 65.42  & 5.55&	85.76&	69.47
				\\
				+Ours-0.7 &6.81 &87.21 &74.36 &6.23 &86.76 &74.01 & 4.24  & 89.31  & 79.70  & 4.59  & 90.51  & 74.27  & 5.47 &	88.45 &	75.59
				\\
				+Ours-0.5 &6.72 &88.76 &77.72 &6.10 &88.34 &77.56 & 4.23  & 90.87  & 83.45 & \underline{4.33*}  & 92.69  & 80.12  & 5.35&	90.17&	79.71
				\\
				+Ours-0.3 &\underline{6.65*} &\underline{91.59*} &\underline{83.36*} &\underline{6.05*} &\underline{91.77*} &\underline{83.50*} & \underline{4.12*}  & \underline{94.01*}  & \underline{90.05*} & 4.35  & \underline{95.38*}  & \underline{88.42*}  & \underline{5.29*}&	\underline{93.19*}&	\underline{86.33*}
				\\
				+Ours-0.1 &\textbf{6.64}&\textbf{91.63} &\textbf{83.60} &\textbf{5.74} &\textbf{91.74} &\textbf{83.55} & \textbf{4.08}  & \textbf{94.74}  & \textbf{90.82} & \textbf{4.28}  & \textbf{96.21}  & \textbf{89.83}  & \textbf{5.19}&	\textbf{93.58}&	\textbf{86.95}
				\\
				\midrule
				\rowcolor[rgb]{ .900,  .900,  .900} Search &6.17&100.00&100.00&5.17&100.00&100.00& 3.83  & 100.00  & 100.00 & 4.38  & 100.00  & 100.00  & 4.89&	100.00&	100.00
				\\
				+Ours-0.9 &6.62 &84.34 &69.09 &5.94 &84.14 &69.31 & 4.18  & 87.96  & 76.84  & 4.49  & 88.55  & 67.93  & 5.31&	86.25&	70.79
				\\
				+Ours-0.7 &6.53 &86.77 &74.34 &5.77 &86.94 &74.46 & 4.18  & 89.52  & 80.79  & 4.37  & 91.78  & 77.67  & 5.21&	88.75&	76.82
				\\
				+Ours-0.5 &6.30 &88.52 &77.57 &5.72 &88.62 &77.46 & \underline{4.15*}  & 91.83  & 85.16  & 4.37  & 93.32  & 82.44  & 5.14&	90.57&	80.66
				\\
				+Ours-0.3 &\underline{6.28*} &\underline{91.45*} &\underline{83.08*} &\underline{5.64*} &\textbf{91.68} &\underline{83.10*} & \textbf{3.96}  & \underline{94.05*}  & \underline{90.79*}  & \underline{4.20*}  & \textbf{95.90}  & \textbf{89.92}  & \underline{5.02*}&	\underline{93.27*}	&\underline{86.72*}
				\\
				+Ours-0.1 &\textbf{6.23} &\textbf{91.54} &\textbf{83.37} &\textbf{5.59} &\underline{91.51*} &\textbf{83.34} & \textbf{3.96}  & \textbf{94.44}  & \textbf{91.84}  & \textbf{4.17}  & \underline{95.77*}  & \underline{89.77*}  & \textbf{4.99}&	\textbf{93.32}&	\textbf{87.08}
				\\
				\midrule[1.2pt]
				
				\multirow{2}[2]{*}{\textbf{Avg.} (Datasets)} &\multicolumn{3}{c|}{TV \cite{KDD2023}} &\multicolumn{3}{c|}{BDD \cite{KDD2023}}& \multicolumn{3}{c|}{AlpacaFarm \cite{INT2023}} & \multicolumn{3}{c||}{FinQA \cite{FinQA2018}} & \multicolumn{3}{c}{\textbf{Avg.}} \\
				\cmidrule{2-16}          & PPL $\downarrow$  & Score $\uparrow$ & ROUGE-L $\uparrow$ &PPL $\downarrow$  & Score $\uparrow$ & ROUGE-L $\uparrow$ &PPL $\downarrow$  & Score $\uparrow$ & ROUGE-L $\uparrow$ & PPL $\downarrow$  & Score $\uparrow$ & ROUGE-2 $\uparrow$ & PPL $\downarrow$  & Score $\uparrow$ & ROUGE $\uparrow$  \\
				\midrule
				\rowcolor[rgb]{ .900,  .900,  .900} {\scriptsize Un-watermarked} &5.93 &	100.00 &	100.00 &	4.67 &	100.00 &	100.00 &	3.49 &	100.00	 &100.00 &	3.71 &	100.00 &	100.00 &	4.45 &	100.00 &	100.00
				\\
				+Ours-0.9 &6.32&	87.93	&75.86&	5.17&	88.31&	77.00&	3.82&	86.31&	73.57&	4.04&	87.01&	63.61&	4.84&	87.39	&72.51
				\\
				+Ours-0.7 &6.24	&90.19&	80.47&	5.09&	90.32&	80.88&	3.79&	89.14&	79.78&	3.97&	90.33&	73.60&	4.77&	89.99&	78.68
				\\
				+Ours-0.5 &6.12	&91.56&	83.15&	5.01&	91.64&	83.50&	3.73&	91.41&	84.69&	3.85&	92.35&	79.86&	4.67&	91.74&	82.80
				\\
				+Ours-0.3 &\underline{6.03*}&	\underline{93.73*}&	\underline{87.48*}&	\underline{4.97*}&	\underline{93.94*}&	\underline{87.72*}&	\underline{3.60*}&	\underline{94.34*}&	\underline{90.91*}&	\underline{3.76*}&	\underline{95.42*}&	\underline{88.83*}&	\underline{4.59*}&	\underline{94.36*}&	\underline{88.73*}
				\\
				+Ours-0.1 &\textbf{5.99}&	\textbf{93.88}&	\textbf{87.85}&	\textbf{4.84}&	\textbf{94.03}&	\textbf{87.97}&	\textbf{3.57}&	\textbf{95.37}&	\textbf{92.67}&	\textbf{3.70}&	\textbf{96.38}&	\textbf{90.90}&	\textbf{4.52}&	\textbf{94.92}&	\textbf{89.85}
				\\
				\bottomrule[1.2pt]
			\end{tabular}%
		}
	\end{table*}%

	The results of Fig. \ref{fig9} and Fig. \ref{fig10} show that the lower the $\eta$: the higher the quality, the shorter the watermark length and payload, and the shorter the embedding time. In addition, we also give some embedding examples, as shown in TABLE \ref{tabf1}.

	\begin{table*}[htbp]
		\centering
		\setlength{\tabcolsep}{1.2mm}
		{\fontsize{8}{9}\selectfont
			\caption{The example of watermarked data under different $\eta$. The watermarked data are given some information about the text at the end, including $\eta$, the watermark length, the Perplexity, the BERTScore, and the embedding time.}\label{tabf1}%
			\begin{tabular}{p{10em}||p{23em}|p{28em}}
				\toprule[1.2pt]
				\multicolumn{1}{c||}{Prompt} & \multicolumn{1}{c|}{Un-watermarked Data} & \multicolumn{1}{c}{Watermarked Data}\\
				\midrule[0.5pt]
				\multicolumn{1}{p{9em}||}{The car wants to get a better position to complete a u-turn, what will the car do next? Please give me a detailed plan.}&The car will first check its surroundings to ensure that it has enough space to complete the u-turn. It will then shift into reverse gear and begin to back up slowly, turning its wheels in the opposite direction of the turn. As the car backs up, it will straighten out its wheels and continue to turn until it is facing in the opposite direction from which it started. Once the car has completed the backward turn, it will shift into drive gear and continue on its way.&
				
				The car will first check its surroundings to ensure that it has enough space to complete a U-turn. It will then shift into reverse gear and begin to back up slowly, turning its wheels in the opposite direction of the turn. As the car backs up, it will straighten its wheels to maintain a steady course. Once the car has backed up far enough, it will turn its wheels in the desired direction and continue to turn until it is facing in the opposite direction of its original path. The car will then shift into drive gear and continue on its journey.
				
				\textbf{($\boldsymbol{\eta = 1.0}$, LLaMA2-7B, Watermark length: 7bits, BERTScore: 92.28, Embedding time: 1.37sec)}\\
				\midrule[0.5pt]
				The car with the camera is driving down an icy road. A dark colored car trys to merge to the left lane. The dark car lost control of direction. The dark car hits on the road side. Could the accident be prevented if the roads are marked clearly?&It is possible that the accident could have been prevented if the roads were marked clearly. When driving on icy roads, it is important to have clear signage and markings to help guide drivers and prevent accidents. If the dark car had seen the clear markings on the road, it may have been able to avoid losing control and hitting the car with the camera. Additionally, clear signage can help drivers slow down and be more cautious when driving on icy roads, which can also help prevent accidents.&It is possible that the accident could have been prevented if the roads were marked clearly. When driving on icy roads, it is important to have clear signage and markings to help guide drivers and prevent accidents. If the dark car had seen the clear markings on the road, it may have been able to avoid losing control and hitting the car with the camera. Additionally, clear signage and markings can help drivers slow down and be more cautious when driving on icy roads, which can also help prevent accidents.
				
				\textbf{($\boldsymbol{\eta = 0.7}$, LLaMA2-7B, Watermark length: 11bits, BERTScore: 99.48, Embedding time: 1.05sec)} \\
				\midrule[0.5pt]
				What year was the Yamato Battleship built?&The Yamato Battleship was built in 1940. \#\#\# Additional Information: The Yamato Battleship was the largest warship ever built by the Japanese Navy, and it served during World War II. It was launched on February 16, 1940, and commissioned on December 16, 1942. The ship was 860 feet long and 128 feet wide, and it had a top speed of 27.5 knots. The Yamato was armed with nine 18.1-inch guns and 12 6-inch guns, and it had a crew of over 2,000 sailors. Despite its size and firepower, the Yamato was sunk by American aircraft during the Battle of Leyte Gulf in October 1944.&The Yamato Battleship was built in 1940. \#\#\# Additional Information:
				The Yamato was the largest battleship ever built by the Imperial Japanese Navy, and it served during World War II. It was launched on February 2, 1940, and commissioned on March 16, 1942. The ship was 860 feet long and 128 feet wide, and it had a top speed of 27.5 knots. The Yamato was armed with nine 18.1-inch guns and 12 6-inch guns, and it had a crew of over 2,000 sailors. Despite its size and firepower, the Yamato was sunk by American aircraft during the Battle of Leyte Gulf in October 1944.
				
				\textbf{($\boldsymbol{\eta = 0.5}$, ChatGLM3-6B, Watermark length: 8bits, BERTScore: 99.11, Embedding time: 1.21sec)}\\
				\midrule[0.5pt]
				The POV car is being rocked as if colliding. The POV car car is rocked. There are black parts flying in front of the car. There are black plastic parts in front of the POV car. What types of vehicles were involved in the accident?&Based on the information provided in the instruction, it appears that there was a collision involving the POV car. The instruction mentions that the POV car is being rocked and that there are black parts flying in front of the car, which suggests that another vehicle may have collided with the POV car. Additionally, the instruction mentions that there are black plastic parts in front of the POV car, which further supports the idea that another vehicle was involved in the accident. Therefore, the types of vehicles involved in the accident are likely to be the POV car and another vehicle.&Based on the information provided in the instruction, it appears that there was a collision involving the POV car. The instruction mentions that the car is being rocked and that black parts and plastic parts are flying in front of it. Additionally, the instruction mentions that there are black plastic parts in front of the POV car, which further supports the idea that another vehicle was involved in the accident. Therefore, the types of vehicles involved in the accident are likely to be the POV car and another vehicle.
				
				\textbf{($\boldsymbol{\eta = 0.3}$, ChatGLM3-6B, Watermark length: 2bits, BERTScore: 95.45, Embedding time: 1.41sec)}\\
				\bottomrule[1.2pt]
			\end{tabular}%
		}
	\end{table*}%

	\subsection{Ablation on Weight $\lambda$ and Number of Read Bits $\varepsilon$}\label{sec542}
\begin{figure}[!htbp]
		\centering
		\includegraphics[width=0.93\linewidth]{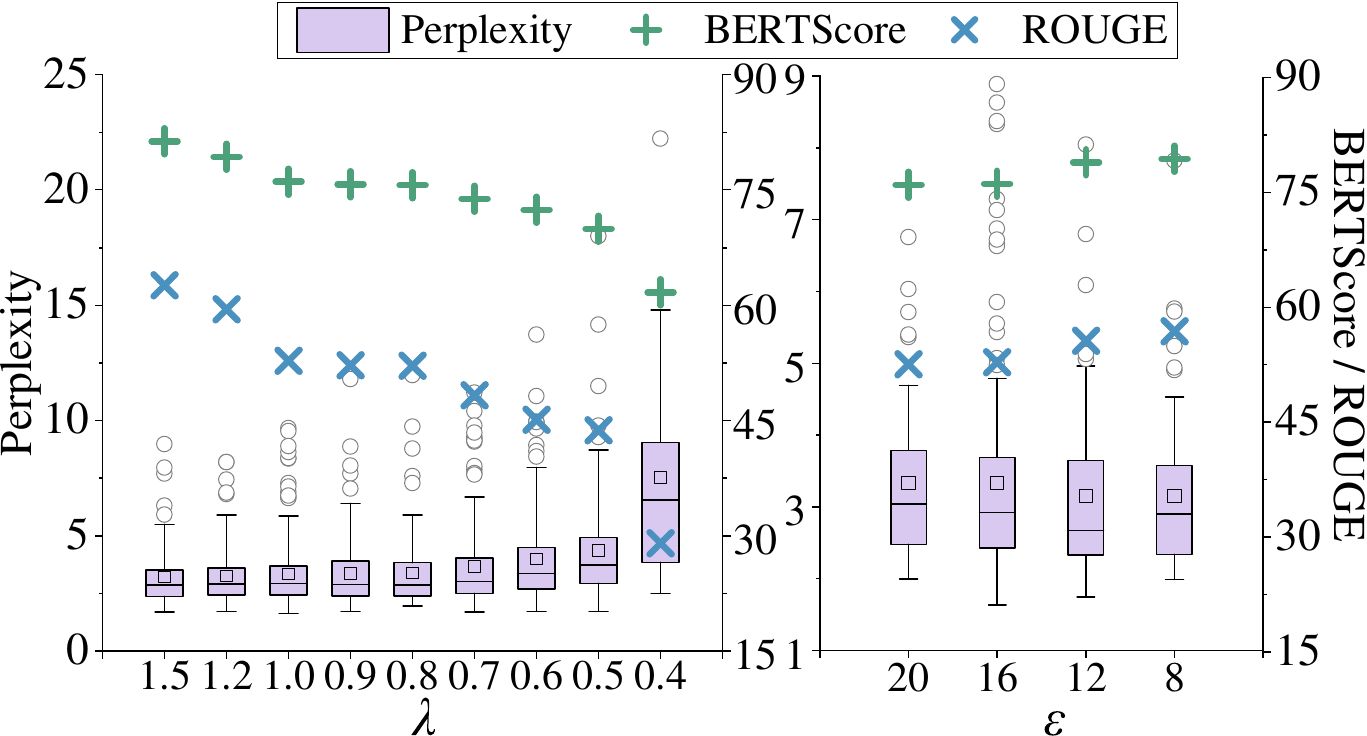}
		\caption{Comparison of data quality under different $\lambda$ (left) and $\varepsilon$ (right).}\label{fig12}
	\end{figure}

	The effects of $\lambda$ and $\varepsilon$ on quality, watermark length, and payload are shown in Fig. \ref{fig12} and TABLE \ref{tab5}.

	\begin{table}[htbp]
		\centering
		\scriptsize
		\setlength{\tabcolsep}{0.45mm}
		\caption{Comparison of watermark length (WL) and payload (P) under different $\lambda$ and $\varepsilon$. The theoretical payload of the CTWL scheme is 0.1 (CTWL-10) and 0.2 (CTWL-5). In the comparison experiment, $\lambda$ and $\varepsilon$ are set to 1.0 and 16 of ITSmark.}\label{tab5}%
		\begin{tabular}{c||ccccccccc|cccc}
			\toprule[1.2pt]
			\multirow{2}[2]{*}{} & \multicolumn{9}{c|}{\textit{$\lambda$}}                                       & \multicolumn{4}{c}{\textit{$\varepsilon$}} \\
			\cmidrule{2-14}          & 1.5   & 1.2   & 1.0     & 0.9   & 0.8   & 0.7   & 0.6   & 0.5   & 0.4   & 20    & 16    & 12    & 8 \\
			\midrule
			WL & 6.69  & 9.50   & 15.82 & 16.66 & 16.58 & 26.27 & 35.05 & 69.42 & 712.96 & 15.79 & 15.82 & 13.46 & 12.05 \\
			P & 0.04  & 0.06  & 0.09  & 0.09  & 0.09  & 0.14  & 0.20   & 0.37  & 2.29  & 0.09  & 0.09  & 0.07  & 0.07 \\
			\bottomrule[1.2pt]
		\end{tabular}%
	\end{table}%

	The results of Fig. \ref{fig12} and TABLE \ref{tab5} show that as $\lambda$ decreases, the overall data quality shows a downward trend, and the watermark length and payload are significantly improved. As $\varepsilon$ decreases, the overall data quality shows an upward trend, and the watermark length and payload are reduced. The impact of changing $\varepsilon$ on data quality and embedding is smaller than that of $\lambda$ and $\eta$ in the previous subsection.

	\subsection{Experimental Summary}\label{sec543}
	In summary, the proposed ITSmark scheme has better data quality than the existing zero-bit and multi-bit watermarking techniques, and the perplexity is reduced by about 20\%. ITSmark can maintain the effect of complete extraction in different datasets, realizing the integrity of copyright. The unforgeability of ITSmark data is enhanced by about 30\% compared with the baseline schemes.
	
	In addition, ITSmark can be verified. When the verification fails, the complete watermark information cannot be extracted using the watermarked data. When the watermark data is tampered with, the tampered location can be accurately traced by simple judgment. 
	
	About the embedding ratio, if the copyright to be embedded is short, or the precision of the watermarked data content is higher, the user can choose a smaller $\eta$ and a larger $\lambda$, which can generate a higher quality watermarked data. If the watermark is long, the user can choose a larger $\eta$ and a smaller $\lambda$, but it also requires suboptimal data quality.

	\section{Discussion \& Future Work}\label{sec7}
	The watermark information embedded in this paper is encoded using ASCII coding, that is, each letter, number, or symbol will be compiled into an 8-bit binary bit stream. When the $\varepsilon$-bit watermark is read once for embedding, the same prefix of the first bit of the determined interval is successfully embedded at the current moment, that is, less than or equal to the $\varepsilon$ bit. This seems to cause part of the watermark information to be lost at the moment of embedding, resulting in a conflict between the 8 bits at the next moment and the real 8 bits, which can be called a segment conflict. However, this segment conflict does not exist. This is because the current successful embedding is less than or equal to the $\varepsilon$ bit, and the beginning of the next moment is the next bit that is successfully embedded. When extracting, the current token can determine a certain interval, thereby determining the current embedded bit less than or equal to the $\varepsilon$ bit, and the beginning to be extracted at the next moment is also the next bit that is successfully embedded. Therefore, the proposed ITSmark is not affected by the reading of the $\varepsilon$ bit value, and there is no segment conflict problem.
	
	Secondly, since the proposed ITSmark is independent of LLM, that is, ITSmark can be grafted onto any LLM. Therefore, if it is to be truly deployed and effective in a wider range of fields, ITSmark can optimize LLMs such as LLaMA with the help of techniques such as Chain of Thought (CoT), or RLHF (Reinforcement Learning from Human Feedback) LLMs that perform well in these fields as base LLMs. The proposed ITSmark scheme can not only add watermark information to data, but also to models such as LLMs, as a watermarking scheme for LLMs to protect the copyright of specific models.
	
	It is worth exploring further that copyright protection in the watermark field may not be established in the AIGC era. In the prisoner model, it is easier for regulators to simulate the sender's content, and it is difficult for the receiver to judge the source of the information, leading to the problem of man-in-the-middle attacks. Although this paper attempts to combine encryption algorithms to alleviate this problem, there is still a lot of room for improvement. Therefore, in the future, we can re-examine the classic attacks on watermarks in the AIGC environment and propose a new paradigm for copyright identification.
	
	Furthermore, as for how to further optimize to narrow the difference between watermarked and un-watermarked. We believe that in terms of white-box models, adaptive or dynamic programming algorithms can be used to make the intervals obtained by tokens more scientific, while effectively embedding watermarks and ensuring the quality of watermarked data. In terms of black-box models, the watermarked data can be controlled from the perspective of indexes such as domain keywords, so that the generation paradigm may get rid of the distortion caused by watermark control data generation.
	
	Finally, since this work focuses on the integrity of ITS data copyright and the security of the verification process, ITSmark is more inclined to the category of fragile watermarks, and “strong robustness” is not the focus of this work. Subsequent research will also study the “ strong robustness” of watermarks, hoping to expand ITSmark's robustness while ensuring its existing functions.

	\section{Conclusion}\label{sec6}
	This paper introduces a watermark scheme for ITS, termed ITSmark. Verification is required before extraction, and only after verification can the watermark be entirely extracted. Extensive experiments show that ITSmark surpasses baselines in data quality, extraction accuracy, and unforgeability. Furthermore, ITSmark also has the necessary functions of ITS, such as permission verification, tampering location tracking, and support for custom embedding location and ratio, which improves the security and reliability of ITS data. We believe that this work greatly protects the copyright of ITS data and provides inspiration for more advanced ITS data security research.

	\section*{Acknowledgement}
	Thanks to the “\textit{IEEE Transactions on Intelligent Transportation Systems}” Editor and Reviewers for their valuable comments!
	
	This work is supported by the National Natural Science Foundation of China (Grant U21B2020) and supported by BUPT Excellent Ph.D. Students Foundation (Grants CX2023120, CX20241055).

	\section*{Contributions}
	Yihao Wang: Conceptualization; Data curation; Investigation; Methodology; Software; original draft. Lingxiao Li: Funding acquisition; Data curation; Investigation; Resources; Software; Validation; original draft. The first two authors contributed equally. Yifan Tang: Formal analysis; Investigation; Resources; Software; Validation; original draft. Ru Zhang: Funding acquisition; Project administration; Supervision; review \& editing. Jianyi Liu: Formal analysis; Supervision; review \& editing.

	\section*{Ethical Statement}
	This research does not involve human participants, personal data, or animal subjects, and as such, does not raise any direct ethical concerns. The data utilized in this study were sourced from publicly available datasets or generated in compliance with applicable laws and regulations.
	
	The methods proposed in this research, including the ITSmark system, are designed to enhance data integrity and security in intelligent transportation systems and do not pose any foreseeable ethical risks when used responsibly.

	
\end{document}